\documentclass[pdftex,twocolumn,epjc3]{svjour3}          

\RequirePackage[T1]{fontenc}

\smartqed  

\RequirePackage{mathptmx}      
\RequirePackage{flushend}
\RequirePackage[numbers,sort&compress]{natbib}
\RequirePackage[colorlinks,citecolor=blue,urlcolor=blue,linkcolor=blue]{hyperref}

\usepackage{latexsym}
\usepackage{amsmath}
\usepackage{amssymb}
\usepackage{latexsym}
\usepackage{amsfonts}
\usepackage{graphicx}
\usepackage{mathrsfs}
\usepackage{hyperref}

\newcommand{\ud}{\mathrm{d}}
\newcommand{\pd}{\partial}

\journalname{Eur. Phys. J. C}

\begin{document}

\title{Polarizations of Gravitational Waves in Horndeski Theory}
\author{Shaoqi Hou\thanksref{e1}
        \and
       Yungui Gong\thanksref{e2}
       \and
       Yunqi Liu\thanksref{e3} 
}

\thankstext{e1}{e-mail: shou1397@hust.edu.cn}
\thankstext{e2}{Corresponding author, e-mail: yggong@hust.edu.cn}
\thankstext{e3}{e-mail: liuyunqi@hust.edu.cn}

\institute{School of Physics, Huazhong University of Science and Technology, Wuhan, Hubei 430074, China}

\date{Accepted: 6 May 2018}

\maketitle

\begin{abstract}
We analyze the polarization content of gravitational waves in Horndeski theory.
Besides the familiar plus and cross polarizations in Einstein's General Relativity, there is one more polarization state which is the mixture of the transverse breathing and longitudinal polarizations.
The additional mode is excited by the massive scalar field.
In the massless limit, the longitudinal polarization disappears, while the breathing one persists.
The upper bound on the graviton mass severely constrains the amplitude of the longitudinal polarization,
which makes its detection highly unlikely by the ground-based or space-borne interferometers in the near future.
However, pulsar timing arrays might be able to detect the polarization excited by the massive scalar field.
Since additional polarization states appear in alternative theories of gravity, the measurement of
the polarizations of gravitational waves can be used to probe the nature of gravity.
In addition to the plus and cross states, the detection of the breathing polarization means that gravitation is mediated by massless spin 2 and spin 0 fields, and
the detection of both the breathing and longitudinal states means that gravitation is propagated by the massless spin 2 and massive spin 0 fields.
\end{abstract} 

\section{Introduction}

The detection of gravitational waves by the Laser Interferometer Gravitational-Wave Observatory (LIGO) Scientific Collaboration and VIRGO Collaboration further supports Einstein's General Relativity (GR) and provides a new tool to study gravitational physics \cite{Abbott:2016blz,Abbott:2016nmj,Abbott:2017vtc,Abbott:2017oio,TheLIGOScientific:2017qsa,Abbott:2017gyy}.
In order to confirm gravitational waves predicted by GR, it is necessary to determine the polarizations of gravitational waves.
This can be done by the network of ground-based Advanced LIGO (aLIGO) and VIRGO,
the future space-borne Laser Interferometer Space Antenna (LISA) \cite{Audley:2017drz} and Tianqin \cite{Luo:2015ght}, and
pulsar timing arrays (e.g., the International Pulsar Timing Array and the European Pulsar Timing Array \cite{Hobbs:2009yy,Kramer:2013kea}).
In fact, in the recent GW170814 \cite{Abbott:2017oio}, the Advanced VIRGO detector joined the two aLIGO detectors,
so they were able to test the polarization content of gravitational waves for the first time.
The result showed that the pure tensor polarizations were favored against pure vector and pure scalar polarizations \cite{Abbott:2017oio,Abbott:2017tlp}.
Additionally, GW170817 is the first observation of a binary neutron star inspiral,
and its electromagnetic counterpart, GRB 170817A, was later observed by the Fermi Gamma-ray Burst Monitor
and the International Gamma-Ray Astrophysics Laboratory \cite{TheLIGOScientific:2017qsa,Goldstein:2017mmi,Savchenko:2017ffs}.
The new era of multi-messenger astrophysics comes.

In GR, the gravitational wave propagates at the speed of light and it has two polarization states, the plus and cross modes.
In alternative metric theories of gravity, there may exist up to six polarizations, so
the detection of the polarizations of gravitational waves can be used to distinguish different theories of gravity and probe the nature of gravity \cite{Isi:2015cva,Isi:2017equ}.
For null plane gravitational waves, the six polarizations are classified by the little group $E(2)$
of the Lorentz group with the help of the six independent Newman-Penrose (NP) variables $\Psi_2$, $\Psi_3$, $\Psi_4$ and $\Phi_{22}$ \cite{Newman:1961qr,Eardley:1974nw,Eardley:1973br}.
In particular, the complex variable $\Psi_4$ denotes the familiar plus and cross modes in GR,
the variable $\Phi_{22}$ denotes the transverse breathing polarization, the complex variable $\Psi_3$ corresponds to the vector-x and vector-y modes,
and the variable $\Psi_2$ corresponds to the longitudinal mode. Under the $E(2)$ transformation, all other modes
can be generated from $\Psi_2$, so if $\Psi_2\neq 0$, then we may see all six modes in some coordinates.
The $E(2)$ classification tells us the general polarization states, but it fails to tell us the correspondence between the polarizations and gravitational theories.
In Brans-Dicke theory \cite{Brans:1961sx}, in addition to the plus and cross modes $\Psi_4$ of the massless gravitons, there exists another breathing mode $\Phi_{22}$ due to the massless Brans-Dicke scalar field \cite{Eardley:1974nw}.

Brans-Dicke theory is a simple extension to GR.
In Brans-Dicke theory, gravitational interaction is mediated by both the metric tensor and the Brans-Dicke scalar field, and the Brans-Dicke field plays the role of Newton's gravitational constant.
In more general scalar-tensor theories of gravity, the scalar field $\phi$ has self-interaction and it is usually massive.
In 1974, Horndeski found the most general scalar-tensor theory of gravity whose action has higher derivatives of $g_{\mu\nu}$ and $\phi$, but the equations of motion are at most the second order \cite{Horndeski:1974wa}.
Even though there are higher derivative terms,
there is no Ostrogradsky instability \cite{Ostrogradsky:1850fid},
so there are three physical degrees of freedom in Horndeski theory, and we expect that there is an extra polarization state in addition to the plus and cross modes.
If the scalar field is massless, then the additional polarization state should be the breathing mode $\Phi_{22}$.

When the interaction between the quantized matter fields and the classical gravitational field is considered, the quadratic terms $R_{\mu\nu\alpha\beta} R^{\mu\nu\alpha\beta}$ and $R^2$ are needed as counterterms to remove the singularities in the energy-momentum tensor \cite{Utiyama:1962sn}.
Although the quadratic gravitational theory is renormalizable \cite{Stelle:1976gc}, the theory has ghost due to the presence of higher derivatives \cite{Stelle:1976gc,Stelle:1977ry}.
However, the general nonlinear $f(R)$ gravity \cite{Buchdahl:1983zz} is free of ghost and it is equivalent to a scalar-tensor theory of gravity \cite{OHanlon:1972xqa,Teyssandier:1983zz}.
The effective mass squared of the equivalent scalar field is $f'(0)/3f''(0)$, and the massive scalar field excites both the longitudinal and transverse breathing modes \cite{Corda:2007hi,Corda:2007nr,Capozziello:2008rq}.
The polarizations of gravitational waves in $f(R)$ gravity were previously discussed in \cite{Capozziello:2010iy,Prasia:2014zya,Alves:2009eg,Alves:2010ms,Rizwana:2016qdq,Myung:2016zdl,Liang:2017ahj}.
The authors in \cite{Alves:2009eg,Alves:2010ms} calculated the NP variables and found that $\Psi_2$, $\Psi_4$ and $\Phi_{22}$ are nonvanishing.
They then claimed that there are at least four polarization states in $f(R)$ gravity.
Recently, it was pointed out that the direct application of the framework of Eardley \textit{et.~al.} (ELLWW framework) \cite{Eardley:1973br,Eardley:1974nw} derived for the null plane gravitational waves to the massive field is not correct, and the polarization state of the equivalent massive field in $f(R)$ gravity
is the mixture of the longitudinal and the transverse breathing modes \cite{Liang:2017ahj}.
Furthermore, the longitudinal polarization is independent of the effective mass of the equivalent scalar field,
so it cannot be used to understand how the polarization reduces to the transverse breathing mode in the massless limit.

Since the polarizations of gravitational waves in alternative theories of gravity are not known in general, 
so we need to study it \cite{Gong:2017bru,Gong:2018cgj,gaogong18}.
In this paper, the focus is on the polarizations of gravitational waves in the most general scalar-tensor theory, Horndeski theory.
It is assumed that matter minimally couples to the metric, so that test particles follow geodesics.
The gravitational wave solutions are obtained from the linearized equations of motion around the flat spacetime background, and the geodesic deviation equations are used to reveal the polarizations of the massive scalar field.
The analysis shows that in Horndeski theory, the massive scalar field excites both breathing and longitudinal polarizations.
The effect of the longitudinal polarization on the geodesic deviation depends on the mass
and it is much smaller than that of the transverse polarization in the aLIGO and LISA frequency bands.
In the massless limit, the longitudinal mode disappears, while the breathing mode persists.

The paper is organized as follows. Section \ref{sec-np} briefly reviews ELLWW framework
for classifying the polarizations of null gravitational waves.
In Section \ref{sec-gwsol}, the linearized equations of motion and the plane gravitational wave solution are obtained.
In Section \ref{sec-pols},  the polarization states of gravitational waves in Horndeski theory are discussed by examining the geodesic deviation equations, and Section \ref{sec-npf} discusses the failure of ELLWW framework for the massive Horndeski theory.
Section \ref{sec-exps} discusses the possible experimental tests of the extra polarizations. In particular,
Section \ref{sec-ligo} mainly calculates the interferometer response functions for aLIGO, and Section \ref{sec-pta} determines the cross-correlation functions for the longitudinal and transverse breathing polarizations.
Finally, this work is briefly summarized in Section \ref{sec-con}.
In this work, we use the natural units and the speed of light in vacuum $c=1$.

\section{Review of ELLWW Framework}\label{sec-np}

Eardley \textit{et.~al.} devised a model-independent framework \cite{Eardley:1974nw,Eardley:1973br} to classify the null gravitational waves in a generic metric theory of gravity based on the Newman-Penrose formalism \cite{Newman:1961qr}. The quasiorthonormal, null tetrad basis $E^\mu_a=(k^\mu,l^\mu,$ $m^\mu,\bar{m}^\mu)$ is chosen to be
\begin{gather}\label{nullb}
  k^\mu=\frac{1}{\sqrt{2}}(1,0,0,1), \\
  l^\mu=\frac{1}{\sqrt{2}}(1,0,0,-1), \\
  m^\mu=\frac{1}{\sqrt{2}}(0,1,i,0), \\
  \bar m^\mu=\frac{1}{\sqrt{2}}(0,1,-i,0),
\end{gather}
where bar  means the complex conjugation. They satisfy the relation $-k^\mu l_\mu=m^\mu\bar m_\mu=1$ and all other inner products vanish.
Since the null gravitational wave propagates in the $+z$ direction, the Riemann tensor is a function of the retarded time $u=t-z$, which implies that $R_{abcd,p}=0$, where $(a,b,c,d)$ range over $(k,l,m,\bar m)$ and $(p,q,r,\cdots)$ range over ($k,m,\bar m$).
The linearized Bianchi identity and the symmetry properties of $R_{abcd}$ imply that
\begin{equation}\label{eq-lbi}
  R_{ab[pq;l]}=R_{ab[pq,l]}  =\frac{1}{3}R_{abpq,l}=0.
\end{equation}
So $R_{abpq}$ is a trivial, nonwavelike constant, and one can set $R_{abpq}=R_{pqab}=0$. One could also verify that $R_{kk}=R_{km}=R_{k\bar m}=R_{mm}=R_{m\bar m}=R_{\bar m\bar m}=0$.
The nonvanishing components of the Riemann tensor are $R_{plql}$.
Since $R_{plql}$ is symmetric in exchanging $p$ and $q$, there are six independent nonvanishing components and they can be expressed in terms of the following four NP variables,
\begin{equation}\label{eq-innp}
\begin{split}
    \Psi_2=&-\frac{1}{6}R_{klkl},\\
    \Psi_3=&-\frac{1}{2}R_{kl\bar ml},\\
    \Psi_4=&-R_{\bar ml\bar ml},\\
    \Phi_{22}=&-R_{ml\bar ml},
\end{split}
\end{equation}
and the remaining nonzero NP variables are $\Phi_{11}=3\Psi_2/2$, $\Phi_{12}=\bar\Phi_{21}=\bar\Psi_3$ and $\Lambda=\Psi_2/2$ \footnote{Note that in Ref.~\cite{Eardley:1974nw},
$\Lambda=-\Psi_2/2$ because Eardley \textit{et.~al.} chose a different sign convention
relating NP variables to the components of the Riemann tensor.}. $\Psi_2$ and $\Phi_{22}$ are real while $\Psi_3$ and $\Psi_4$ are complex.
So there are exactly six real independent variables, and they can be used to replace the six components of the electric part of the Riemann tensor.

These four NP variables $\{\Psi_2,\Psi_3,\Psi_4,\Phi_{22}\}$
can be classified according to their transformation properties under the group $E(2)$,
the little group of the Lorentz group for massless particles.
$E(2)$ consists of local transformations preserving the null vector $k^\mu$, i.e., under $E(2)$ transformation,
\begin{equation}\label{eq-e2tnp}
  \begin{split}
    \Psi'_2=&\Psi_2,\\
    \Psi'_3=&e^{-i\vartheta}(\Psi_3+3\bar\rho\Psi_2),\\
    \Psi'_4=&e^{-i2\vartheta}(\Psi_4+4\bar\rho\Psi_3+6\bar\rho^2\Psi_2),\\
    \Phi'_{22}=&\Phi_{22}+2\rho\Psi_3+2\bar\rho\bar\Psi_3+6\rho\bar\rho\Psi_2,
  \end{split}
\end{equation}
where $\vartheta\in[0,2\pi)$ represents a rotation around the $z$ direction and the complex number $\rho$ denotes a translation in the Euclidean 2-plane.
From these transformations, one finds out that only $\Psi_2$ is invariant and the amplitudes of the four NP variables are not observer-independent.
However, the absence (zero amplitude) of some of the four NP variables is observer-independent.
Based on this, six classes are defined as follows \cite{Eardley:1974nw}.
\begin{description}
  \item[Class II$_6$] $\Psi_2\ne0$. All observers measure the same nonzero amplitude of the $\Psi_2$ mode,
  but the presence or absence of all other modes is observer-dependent.
  \item[Class III$_5$] $\Psi_2=0$, $\Psi_3\ne0$. All observers measure the absence of the $\Psi_2$ mode and the presence of the $\Psi_3$ mode, but
  the presence or absence of $\Psi_4$ and $\Phi_{22}$ is observer-dependent.
  \item[Class N$_3$] $\Psi_2=\Psi_3=0,\,\Psi_4\ne0\ne\Phi_{22}$. The presence or absence of all modes is observer-independent.
  \item[Class N$_2$] $\Psi_2=\Psi_3=\Phi_{22}=0,\,\Psi_4\ne0$. The presence or absence of all modes is observer-independent.
  \item[Class O$_1$] $\Psi_2=\Psi_3=\Psi_4=0,\,\Phi_{22}\ne0$. The presence or absence of all modes is observer-independent.
  \item[Class O$_0$] $\Psi_2=\Psi_3=\Psi_4=\Phi_{22}=0$. No wave is observed.
\end{description}
Apparently, Class II$_6$ is the most general one. If $\Psi_2\neq 0$, then we may see all six modes in some coordinates. By setting successive variables $\{\Psi_2,\Psi_3,\Psi_4,\Phi_{22}\}$ to zero, one obtains less and less general classes, and eventually, Class O$_0$ which is trivial and represents no wave.
These NP variables can also be grouped according to their helicities.
By setting $\rho=0$ in Eq.~\eqref{eq-e2tnp}, one sees that $\Psi_2$ and $\Phi_{22}$ have helicity 0, $\Psi_3$ has helicity 1 and $\Psi_4$ has helicity 2.

In order to establish the relation between $\{\Psi_2,\Psi_3,\Psi_4,\Phi_{22}\}$ and the polarizations of the gravitational wave,
one needs to examine the geodesic deviation equation in the Cartesian coordinates \cite{Eardley:1974nw},
\begin{equation}\label{geodevl}
  \ddot x^j=\frac{d^2x^j}{dt^2}=-R_{tjtk}x^k,
\end{equation}
where $x^j$ represents the deviation vector between two nearby geodesics and $j,\,k=1,2,3$.
The so-called electric part $R_{tjtk}$ of the Riemann tensor is given by the following matrix,
\begin{equation}\label{eq-rtjtkm}
  R_{tjtk}=\left(
  \begin{array}{ccc}
    -\frac{1}{2}(\Re\Psi_4+\Phi_{22}) & \frac{1}{2}\Im\Psi_4 & -2\Re\Psi_3 \\
    \frac{1}{2}\Im\Psi_4 & \frac{1}{2}(\Re\Psi_4-\Phi_{22}) & 2\Im\Psi_3 \\
    -2\Re\Psi_3 & 2\Im\Psi_3 & -6\Psi_2
  \end{array}
  \right),
\end{equation}
where $\Re$ and $\Im$ stand for the real and imaginary parts. So these NP variables can be used to label the polarizations of null gravitational waves.
More specifically, $\Re\Psi_4$ and $\Im\Psi_4$ represent the plus and the cross polarizations, respectively;
$\Phi_{22}$ represents the transverse breathing polarization, and $\Psi_2$ represents the longitudinal polarization; finally,
$\Re\Psi_3$ and $\Im\Psi_3$ represent vector-$x$ and vector-$y$ polarizations, respectively.
FIG.~1 in Ref.~\cite{Eardley:1974nw} displays how these polarizations deform a sphere of test particles, which will not be reproduced here.
In terms of $R_{tjtk}$, the plus mode is specified by $\hat{P}_+=-R_{txtx}+R_{tyty}$, the cross mode is specified by $\hat{P}_\times=R_{txty}$, the transverse
breathing mode is represented by $\hat{P}_b=R_{txtx}+R_{tyty}$, the vector-$x$ mode is represented by $\hat{P}_{xz}=R_{txtz}$,
the vector-$y$ mode is represented by $\hat{P}_{yz}=R_{tytz}$,
and the longitudinal mode is represented by $\hat{P}_l=R_{tztz}$.
For null gravitational waves, the four NP variables $\{\Psi_2,\Psi_3,\Psi_4,\Phi_{22}\}$ with six
independent components are related with the six electric components of Riemann tensor,
and they provide the six independent polarizations $\{\hat{P}_+,\hat{P}_\times,\hat{P}_b,\hat{P}_{xz},\hat{P}_{yz},\hat{P}_l\}$.
According to the $E(2)$ classification, the longitudinal mode which corresponds to nonzero $\Psi_2$ belongs to the most general class $\Pi_6$.
The presence of the longitudinal mode means that all six polarizations are present in some coordinate systems.
Although we obtain the six polarizations according to the $E(2)$ classification for general metric theory of gravity,
the polarizations of gravitational waves in alternative theories of gravity
are unknown in general, so we need to discuss them case by case.

It is now ready to apply this framework to discuss some specific metric theories of gravity.
For example, GR predicts the existence of the plus and the cross polarizations, and it can be easily checked that only $\Psi_4\ne0$.
For Brans-Dicke theory \cite{Brans:1961sx}, there is one more polarization, the transverse breathing polarization, excited by the \emph{massless} scalar field, so in addition to $\Psi_4\ne0$, $\Phi_{22}$ is also nonzero \cite{Eardley:1974nw}.
In particular, $\Psi_2=0$, so the longitudinal polarization is absent in Brans-Dicke theory. So for Brans-Dicke theory,
\begin{equation}\label{eq-rtjtkm-bd}
  R_{tjtk}^\mathrm{BD}=\left(
  \begin{array}{ccc}
    -\frac{1}{2}(\Re\Psi_4+\Phi_{22}) & \frac{1}{2}\Im\Psi_4 & 0 \\
    \frac{1}{2}\Im\Psi_4 & \frac{1}{2}(\Re\Psi_4-\Phi_{22}) & 0 \\
    0 & 0 & 0
  \end{array}
  \right).
\end{equation}

In the next section, the plane gravitational wave solution to the linearized equation of motion will be obtained for Horndeski theory, and the polarization content will be determined.
It will show that because of the massive scalar field,
the electric part of the Riemann tensor takes a different form from Eq.~\eqref{eq-rtjtkm-bd}.
Its components are expressed in terms of a different set of NP variables.

\section{Gravitational Wave Polarizations in Horndeski Theory}\label{sec-gwsol}

In this section, the polarization content of the plane gravitational wave in Horndeski theory will be determined. The action of Horndeski theory is \cite{Horndeski:1974wa},
\begin{equation}
\label{acth}
  S=\int d^4x\sqrt{-g}(L_2+L_3+L_4+L_5),
\end{equation}
where
\begin{gather*}
L_2=K(\phi,X),\quad L_3=-G_3(\phi,X)\Box \phi, \\ L_4=G_4(\phi,X)R+G_{4,X}\left[(\Box\phi)^2-(\nabla_\mu\nabla_\nu\phi)(\nabla^\mu\nabla^\nu\phi)\right], \\
L_5=G_5(\phi,X)G_{\mu\nu}\nabla^\mu\nabla^\nu\phi-\frac{1}{6}G_{5,X}[(\Box\phi)^3\\
-3(\Box\phi)(\nabla_\mu\nabla_\nu\phi)(\nabla^\mu\nabla^\nu\phi)\\
+2(\nabla^\mu\nabla_\alpha\phi)(\nabla^\alpha\nabla_\beta\phi)(\nabla^\beta\nabla_\mu\phi)].
\end{gather*}
Here, $X=-\nabla_\mu\phi\nabla^\mu\phi/2$, $\Box\phi=\nabla_\mu\nabla^\mu\phi$,
the functions $K$, $G_3$, $G_4$ and $G_5$ are arbitrary functions of $\phi$ and $X$, and $G_{j,X}(\phi,$ $X)=\partial G_j(\phi,X)/\partial X$ with $j=4,5$.
Horndeski theory reduces to several interesting theories as its subclasses by suitable choices of these functions. For example, one obtains GR with $K=G_3=G_5=0$ and $G_4=1/(16\pi G)$.
For Brans-Dicke theory, $G_3=G_5=0$, $K=2\omega X/\phi$ and $G_4=\phi$.
And finally, to reproduce $f(R)$ gravity, one can set $G_3=G_5=0$, $K=f(\phi)-\phi f'(\phi)$ and $G_4=f'(\phi)$ with $f'(\phi)=d f(\phi)/d\phi$.

The variational principle gives rise to the equations of motion.
The full set of equations can be found in Ref.~\cite{Kobayashi:2011nu}, for instance.
For the purpose of this work, one expands the metric tensor field $g_{\mu\nu}$ and $\phi$ in the following way,
\begin{gather}
  g_{\mu\nu}=\eta_{\mu\nu}+h_{\mu\nu}, \\
  \phi=\phi_0+\varphi,
\end{gather}
where $\phi_0$ is a constant.
The equations of motion are expanded up to the linear order in $h_{\mu\nu}$ and $\varphi$,
\begin{gather}
   -\frac{1}{2}K(0)+ G_{4}(0)G_{\mu\nu}^{(1)}-G_{4,\phi}(0)(\varphi_{\mu\nu}-\eta_{\mu\nu}\Box\varphi) =0,\label{eq-einhgb}\\
    K_{,\phi}(0)+[K_{,X}(0)-2G_{3,\phi}(0)]\Box\varphi+K_{,\phi\phi}(0)\varphi\nonumber\\
     +G_{4,\phi}(0)R^{(1)}= 0,\label{eq-schgb}
\end{gather}
where $\Box=\pd^\mu\pd_\mu$ from now on, $G_4(0)=G_4(\phi_0,0)$, $K(0)=K(\phi_0,0)$, and $G_{\mu\nu}^{(1)}$ and $R^{(1)}$ are the linearized Einstein tensor and Ricci scalar, respectively. In addition,
the symbol $K_{,X}(0)$ means the value of $K_{,X}=\partial K/\partial X$ at $\phi=\phi_0$ and $X=X_0=0$, and the remaining symbols can be understood similarly.

In order to obtain the gravitational wave solutions around the flat background,
one requires that $g_{\mu\nu}=\eta_{\mu\nu}$ and $\phi=\phi_0$ be the solution to Eqs.~(\ref{eq-einhgb}) and (\ref{eq-schgb}), which implies that
\begin{equation}\label{eq-flatcon}
  K(0)=0,\quad K_{,\phi}(0)=0.
\end{equation}
Substituting this result into Eqs.~(\ref{eq-einhgb}) and (\ref{eq-schgb}), and after some algebraic manipulations, one gets,
\begin{gather}
\label{eq-dcphgseq}
(\Box-m^2)\varphi = 0,\\
\label{eq-einhga}
G_{\mu\nu}^{(1)}-\frac{G_{4,\phi}(0)}{G_4(0)}(\partial_\mu\partial_\nu\varphi-\eta_{\mu\nu}\Box\varphi)=0,
\end{gather}
where $G_4(0)\neq 0$ and the mass squared of the scalar field is
\begin{equation}
\label{msq}
m^2=-\frac{K_{,\phi\phi}(0)}{K_{,X}(0)-2G_{3,\phi}(0)+3G_{4,\phi}^2(0)/G_{4}(0)}.
\end{equation}
For Brans-Dicke theory, $K_{,\phi\phi}(0)=0$, so $m=0$. For the $f(R)$ theory, one gets $m^2=f'(0)/3f''(0)$, which agrees with the previous work \cite{Liang:2017ahj}.

One can decouple Eq.~(\ref{eq-einhga}) from Eq.~\eqref{eq-dcphgseq} by reexpressing them in terms of the auxiliary tensor field $\tilde h_{\mu\nu}$ defined by,
\begin{equation}
\label{auht}
\tilde h_{\mu\nu}=h_{\mu\nu}-\frac{1}{2}\eta_{\mu\nu}\eta^{\alpha\beta}h_{\alpha\beta}-\eta_{\mu\nu}\sigma\varphi,
\end{equation}
with $\sigma=G_{4,\phi}(0)/G_{4}(0)$.
The original metric tensor perturbation can be obtained by inverting the above relation, i.e.,
\begin{equation}\label{hinht}
 h_{\mu\nu}=\tilde h_{\mu\nu}-\frac{1}{2}\eta_{\mu\nu}\eta^{\alpha\beta}\tilde h_{\alpha\beta}-\eta_{\mu\nu}\sigma\varphi.
\end{equation}
Therefore, Eq.~(\ref{eq-einhga}) becomes
\begin{equation}\label{eq-dcphgeineq}
  \pd_\rho\pd_{(\mu}\tilde h^\rho_{\nu)}-\frac{1}{2}\Box\tilde h_{\mu\nu}-\frac{1}{2}\eta_{\mu\nu}\pd_\rho\pd_\sigma\tilde h^{\rho\sigma} =0,
\end{equation}

The equations of motion remain invariant under the gauge transformation,
\begin{equation}\label{gatrans}
  \varphi' = \varphi,\quad\tilde h'_{\mu\nu} = \tilde h_{\mu\nu}-\pd_\mu\xi_\nu-\pd_\nu\xi_\mu+\eta_{\mu\nu}\pd_\rho\xi^\rho,
\end{equation}
with $x'^\mu=x^\mu+\xi^\mu$. Therefore, one can choose the transverse traceless gauge $\partial_\nu \tilde h^{\mu\nu}=0$, $\eta^{\mu\nu}\tilde h_{\mu\nu}=0$ by using the freedom of coordinate transformation, and the linearized Eqs. \eqref{eq-dcphgseq} and \eqref{eq-einhga} become the wave equations
\begin{gather}
\label{eq-sceqf}
(\Box-m^2)\varphi = 0,\\
\label{eq-eineqf}
\Box\tilde h_{\mu\nu} = 0.
\end{gather}
By the analogue of $\bar h_{\mu\nu}=h_{\mu\nu}-\eta_{\mu\nu}\eta^{\rho\sigma}h_{\rho\sigma}$ in GR,
it is easy to see that the field $\tilde{h}_{\mu\nu}$ represents the familiar massless graviton and it has two polarization states:
the plus and cross polarizations. The scalar field $\varphi$ is massive in general,
and it decouples from the massless tensor field $\tilde{h}_{\mu\nu}$.
Suppose the massless and the massive modes both propagate in the $+z$ direction with the wave vectors,
\begin{equation}
\label{wvs}
  k^\mu=(\Omega,0,0,\Omega),\quad q^\mu=(q_t,0,0,q_z),
\end{equation}
respectively, where the dispersion relation for the scalar field is  $q_t^2-q_z^2=m^2$.
The propagation speed of the massive scalar field is $v=\sqrt{q_t^2-m^2}/q_t$. The plane wave solutions to Eqs. \eqref{eq-sceqf} and \eqref{eq-eineqf} take the following form
\begin{gather}
\tilde h_{\mu\nu} = e_{\mu\nu}e^{-ik\cdot x},\label{wavesol-h} \\
\varphi = \varphi_0e^{-iq\cdot x},\label{wavesol-s}
\end{gather}
where $\varphi_0$ and $e_{\mu\nu}$ are the amplitudes of the waves with $k^\nu e_{\nu\mu}=0$ and $\eta^{\mu\nu}e_{\mu\nu}=0$.

\subsection{Polarizations}\label{sec-pols}

The polarizations of gravitational wave can be extracted by studying the relative acceleration of two nearby test particles moving in the field of the gravitational wave.
One assumes that the matter fields  minimally couple with the metric tensor $g_{\mu\nu}$, while there are no direct interactions between the matter fields and the scalar field $\phi$.
Therefore, freely falling test particles follow geodesics, and the relative acceleration is given by the linearized geodesic deviation equations Eq.~\eqref{geodevl}.
The polarizations of gravitational wave can be understood by placing a sphere of test particles in the spacetime,
and studying how this sphere deforms.
With the solutions \eqref{wavesol-h} and \eqref{wavesol-s}, one calculates the electric part $R_{tjtk}$ for the plane wave solution \eqref{wavesol-h} and \eqref{wavesol-s}.
Written as a $3\times3$ matrix, it is given by
\begin{equation}\label{rtjtk}
\begin{split}
  &R_{tjtk}=\\
  &\left(
  \begin{array}{ccc}
    -\frac{1}{2}q_t^2\sigma\varphi+\frac{1}{2}\Omega^2\tilde h_{xx} & \frac{1}{2}\Omega^2\tilde h_{xy} & 0 \\
    \frac{1}{2}\Omega^2\tilde h_{xy} & -\frac{1}{2}q_t^2\sigma\varphi-\frac{1}{2}\Omega^2\tilde h_{xx} & 0 \\
    0 & 0 & -\frac{1}{2}m^2\sigma\varphi
  \end{array}\right).
\end{split}
\end{equation}
From this expression, one can easily recognize the familiar plus and cross polarizations by setting $\varphi=0$,
which leads to a symmetric, traceless matrix with nonvanishing components $R_{txtx}=-R_{tyty}=\Omega^2\tilde h_{xx}/2$ and $R_{txty}=R_{tytx}=\Omega^2\tilde h_{xy}/2$.
Indeed, $\tilde h_{\mu\nu}$ generates the plus and the cross polarizations.

To study the polarizations caused by the scalar field, one sets $\tilde h_{\mu\nu}=0$.
If the scalar field is massless ($m=0$) as in Brans-Dicke theory,
one finds out that $R_{tztz}=0$, and the scalar field excites only the transverse breathing polarization with $R_{txtx}=R_{tyty}$.
If the scalar field is massive, one can perform a Lorentz boost such that $q_z=0$, i.e., one works in the rest frame of the scalar field.
In the rest frame, from Eq. \eqref{rtjtk}, we find that $q_t^2=m^2$ and $R_{txtx}=R_{tyty}=R_{tztz}\neq 0$.
the geodesic deviation equations are,
\begin{equation}\label{geodevsl}
  \ddot x^j=\frac{1}{2}m^2\sigma\varphi x^j,\, j=1,2,3.
\end{equation}
Suppose the initial deviation vector between two geodesics is $x^j_{0}=(x_0,y_0,z_0)$, and one integrates the above equations twice to obtain the changes in the deviation vector,
\begin{equation}\label{chcos}
  \delta x^j\approx -\frac{1}{2}\sigma \varphi x^j_{0}.
\end{equation}
From the above equation, one finds that if $\sigma$ is independent of the mass $m$, so are $\delta x^j$.
However, for $f(R)$ gravity, $m^2\sigma=1/3$, so $\delta x^j$ is proportional to $1/m^2$ \cite{Liang:2017ahj}.
Eq.~\eqref{chcos} implies that the sphere of test particles will oscillate isotropically in all directions,
so the massive scalar field excites the longitudinal mode in addition to the breathing mode.
Note that in the rest frame of a massive field (such as $\varphi$) one cannot take the massless limit, as there is no rest frame for a massless field
with the speed of light as its propagation speed.
However, the massless limit can be taken in Eq. \eqref{rtjtk} if the rest frame condition $q_t=m$ is not imposed a prior.
In the massless limit, $R_{tztz}=0$ and $R_{txtx}=R_{tyty}\neq 0$.

\begin{figure}
  \centering
  \includegraphics[width=0.3\textwidth]{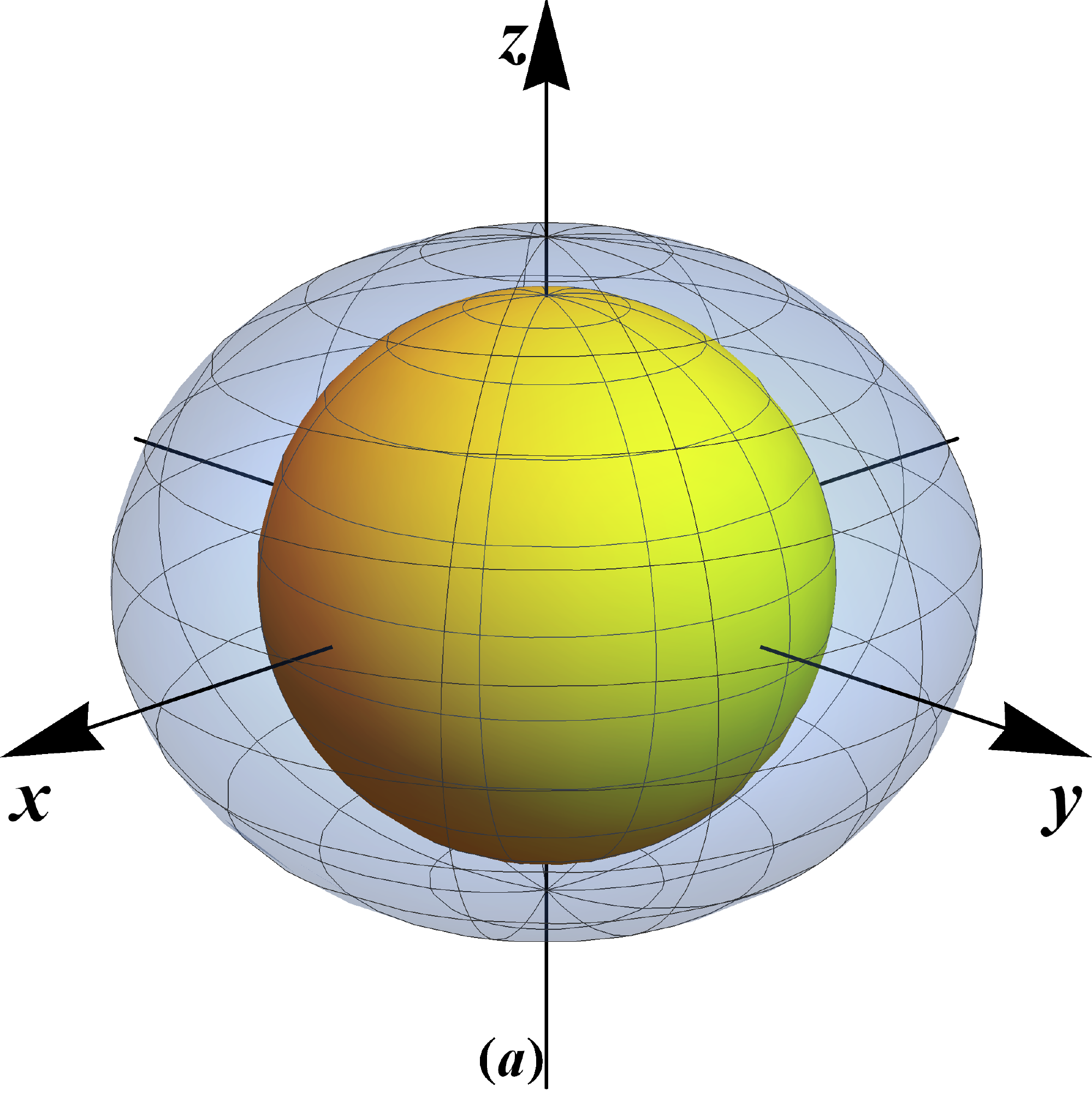}
  \includegraphics[width=0.3\textwidth]{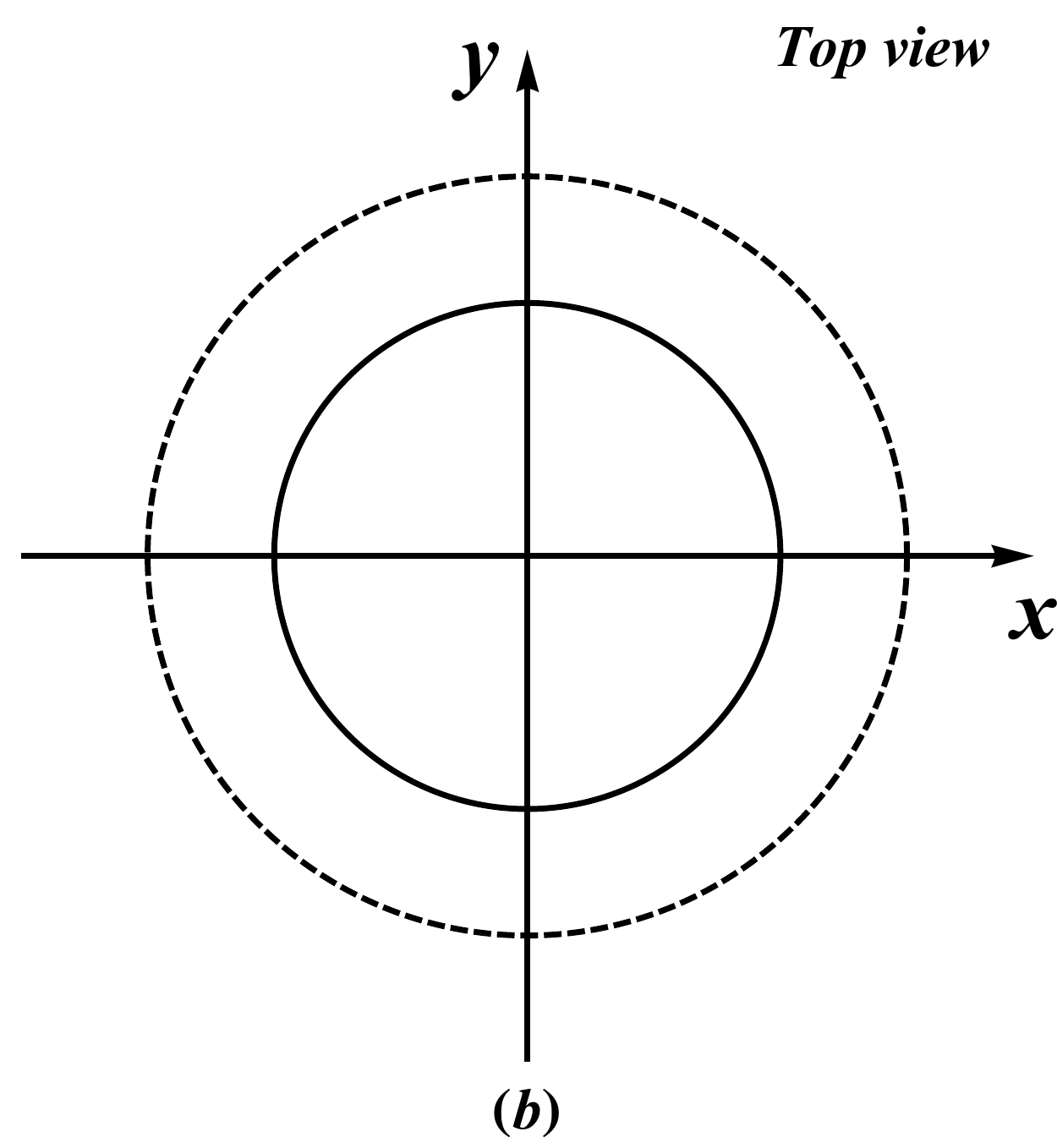}
  \includegraphics[width=0.3\textwidth]{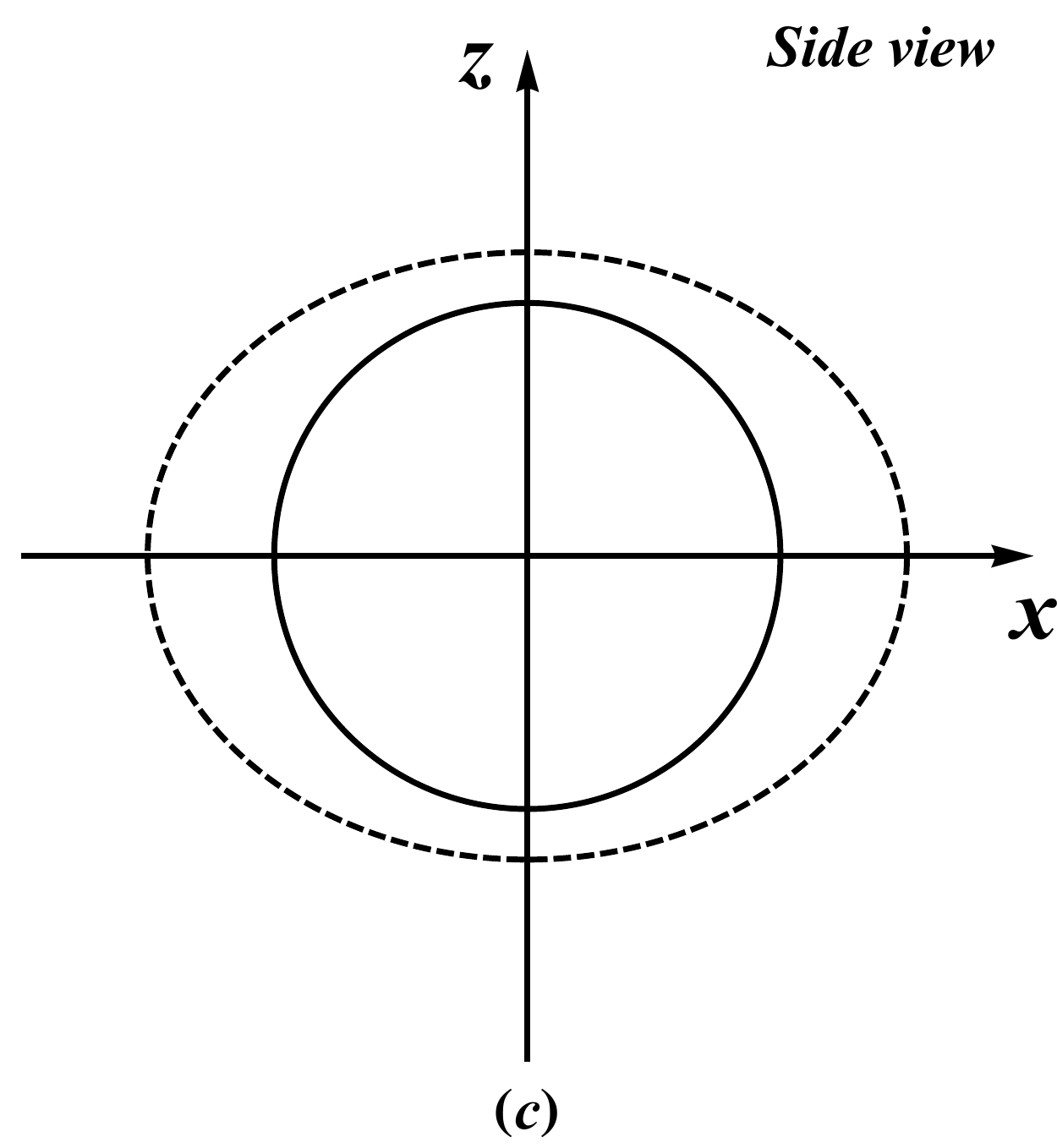}
  \caption{The transverse breathing and the longitudinal polarizations excited by the massive scalar field.}\label{fig-scpolmassive}
\end{figure}

\begin{figure}
  \centering
  \includegraphics[width=0.3\textwidth]{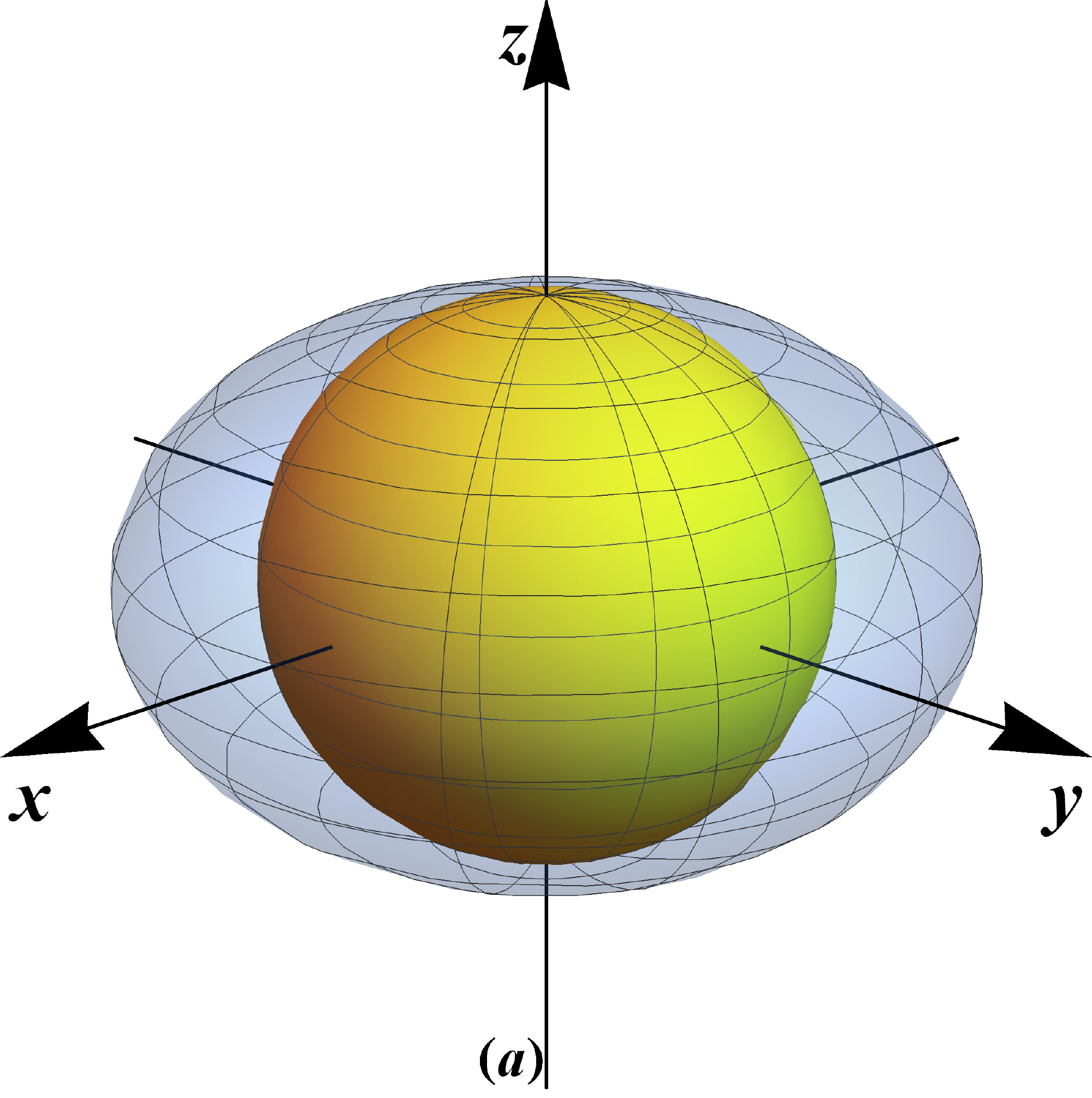}
  \includegraphics[width=0.3\textwidth]{breathingMode.pdf}
  \includegraphics[width=0.3\textwidth]{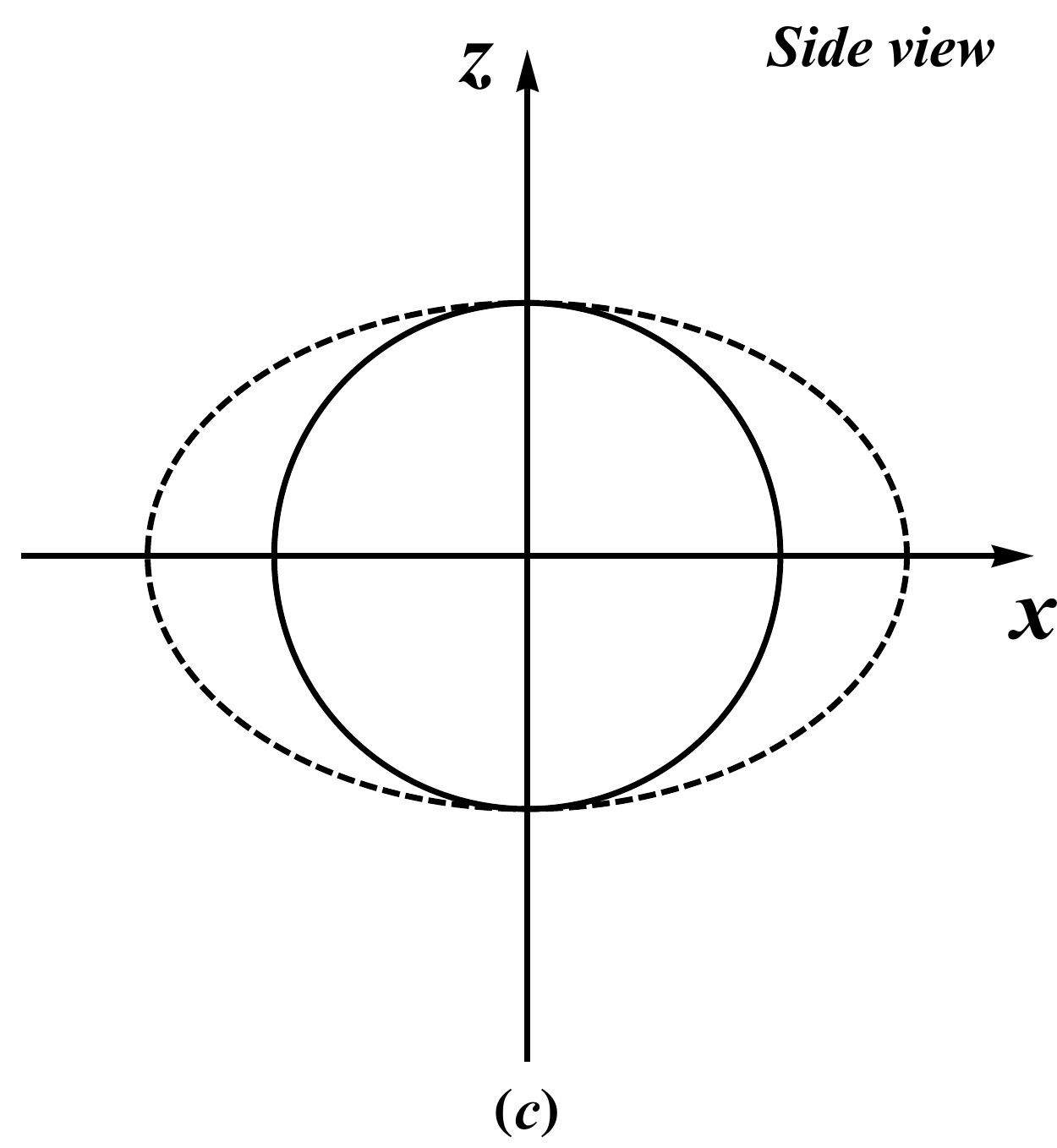}
  \caption{The transverse breathing polarization excited by the massless scalar field.}\label{fig-scpolmassless}
\end{figure}

However, in the actual observation, it is highly unlikely that the test particles, such the mirrors in aLIGO/VIRGO, are at the rest frame of the (massive) scalar gravitational wave.
So one should also study how the scalar gravitational wave deforms the sphere when $q_z\ne0$.
In this case, the deviation vector is given by
\begin{gather}
  \delta x\approx -\frac{1}{2}\sigma \varphi x_0,\label{chcos-x}\\
  \delta y \approx -\frac{1}{2}\sigma \varphi y_0,\label{chcos-y}\\
  \delta z\approx -\frac{1}{2}\frac{m^2}{q_t^2}\sigma\varphi z_0.\label{chcos-z}
\end{gather}
When the gravitational wave passes by, this sphere deforms as shown in Figures \ref{fig-scpolmassive} and \ref{fig-scpolmassless}
for the massive and massless scalars, respectively. In each figure, Panel $(a)$ shows how the sphere deforms in the 3 dimensional
space in which the semitransparent ellipsoids are the results of deforming the spheres, and Panels $(b)$ and $(c)$ show the top
view and the side view  with the dashed ellipses representing the results of the deformation of the circles, respectively.
Because of the cylindrical symmetry around the $z$ axis, the $xOy$ cross section is not displayed in each figure.
In both figures, one finds out that Panels $(b)$ behave the same, as Eqs. \eqref{chcos-x} and \eqref{chcos-y}
are independent of the mass. So Panels $(b)$ represent the transverse breathing polarization,
as the circle (the intersection of the original sphere with the $xOy$ plane) deforms
to a circle of a different size. From Panel $(c)$ in Figure \ref{fig-scpolmassive},
one finds out that the $z$ coordinates of the test particles change (except the test particles on $xOy$ plane),
and the change in $z$ is smaller than that in $x$ (or $y$). This panel represents the longitudinal polarization.
In contrast,  Panel $(c)$ in Figure \ref{fig-scpolmassless} shows that the $z$ coordinates of the test particles remain the same,
so in the massless case, there does not exist the longitudinal polarization.
Note that in the above discussion, we distinguish the breathing polarization (described by Eqs.~\eqref{chcos-x} and \eqref{chcos-y}) from the longitudinal one (described by Eq.~\eqref{chcos-z})
in the massive case,
but since they are excited by the same field $\varphi$, they represent a single degree of freedom.
We thus call the polarization state excited by the scalar a mix polarization of the breathing and the longitudinal polarizations.

In summary, the polarizations of gravitational waves in Horndeski theory
include the plus and cross polarizations induced by the spin 2 field $\tilde h_{\mu\nu}$.
The scalar field excites both the transverse breathing and longitudinal polarizations if it is massive.
However, if it is massless, the scalar field excites merely the transverse breathing polarization.
In terms of the basis introduced for the massless fields in \cite{Eardley:1974nw},
there are three polarizations: the plus state $\hat{P}_+$, the cross state $\hat{P}_\times$ and the mix state of $\hat{P}_b$ and $\hat{P}_l$.
In the massless limit, the mix state reduces to the pure state $\hat{P}_b$.
Note that the vector modes $\hat{P}_{xz}$ and $\hat{P}_{yz}$ are absent in Horndeski theory, this seems to be in conflict with
the $E(2)$ classification because the longitudinal mode $\hat{P}_l$ is present,
so we need to discuss the application of $E(2)$ classification.

\subsection{Newman-Penrose variables}\label{sec-npf}

Now we calculate the NP variables with the plane wave solution \eqref{wavesol-h} and \eqref{wavesol-s}, we get
\begin{equation}
\label{psi20}
\begin{split}
\Psi_2=&\frac{1}{12}(R_{txtx}+R_{tyty}-2R_{tztz}\\
&+2R_{xyxy}-R_{xzxz}-R_{yzyz})+\frac{1}{2}iR_{tzxy}\\
=&0,
\end{split}
\end{equation}
and the nonvanishing ones are
\begin{gather}
 \Psi_4=-\Omega^2(\tilde h_{xx}-i\tilde h_{xy}),\label{nps-psi4}\\
 \Phi_{22}=\frac{(q_t+\sqrt{q_t^2-m^2})^2}{4}\sigma\varphi,\label{nps-phi22}\\
 \Phi_{00}=\frac{(q_t-\sqrt{q_t^2-m^2})^2}{(q_t+\sqrt{q_t^2-m^2})^2}\Phi_{22},\label{nps-phi00}\\
 \Phi_{11}=\Lambda=\frac{m^2}{2(q_t+\sqrt{q_t^2-m^2})^2}\Phi_{22}.\label{nps-phi11-lambda}
\end{gather}
Note that $\Psi_2=-R_{tztz}/6$ for null gravitational waves only, in general case we should use Eq. \eqref{psi20}.
The naive application of the ELLWW framework to the massive case gives the conclusion that Horndeski gravity has the plus,
cross and transverse breathing polarizations, because $\Psi_2=\Psi_3=0$.
Moreover, $\Phi_{00}$ is absent and both $\Phi_{11}$ and $\Lambda$ are proportional to $\Psi_2$ in the ELLWW framework, which is in contradiction with Eqs.~\eqref{nps-phi00} and Eqs.~\eqref{nps-phi11-lambda}.
These indicate the failure of this framework for the massive Horndeski theory.
In particular, the absence of $\Psi_2$ means that there would be no longitudinal polarization if the ELLWW framework were correct.
However, the previous discussion in Section \ref{sec-pols} clearly shows the existence of the longitudinal polarization.
Therefore, the ELLWW framework cannot be applied to a
theory which predicts the existence of massive gravitational waves.
The massive scalar field excites both the breathing and longitudinal polarizations.

In fact, $R_{tjtk}$ can be rewritten in terms of NP variables as a matrix displayed below,
\begin{equation}\label{rtjtknp}
  R_{tjtk}= \left(
  \begin{array}{ccc}
    \Upsilon-\frac{1}{2}\Re\Psi_4 & \frac{1}{2}\Im\Psi_4 & 0 \\
    \frac{1}{2}\Im\Psi_4 & \Upsilon+\frac{1}{2}\Re\Psi_4 & 0 \\
    0 & 0 & -2(\Lambda+\Phi_{11})
  \end{array}\right),
\end{equation}
with $\Upsilon=-2\Lambda-\frac{\Phi_{00}+\Phi_{22}}{2}$.
It is rather different from Eq.~\eqref{eq-rtjtkm-bd}. From the above expression,
it is clear that some linear combinations of $\Phi_{00}$, $\Phi_{11}$, $\Phi_{22}$ and $\Lambda$ correspond to the transverse breathing and the longitudinal polarizations.
In particular, $\Upsilon=-2\Lambda-(\Phi_{00}+\Phi_{22})/2$ represents the transverse breathing polarization,
while $-2(\Lambda+\Phi_{11})$ represents the longitudinal polarization.
The plus and cross polarizations are still represented by $\Re\Psi_4$ and $\Im\Psi_4$, respectively.
In the massless limit, $\Phi_{00}=\Phi_{11}=\Lambda=0$ according to Eqs. \eqref{nps-phi00} and \eqref{nps-phi11-lambda},
so Eq. \eqref{rtjtknp} takes the same form as Eq. \eqref{eq-rtjtkm-bd} since the ELLWW framework applies in this case.
Note that when the scalar field is massive, Eq.~\eqref{rtjtknp} actually takes the same form as Eq.~\eqref{rtjtk} by setting $q_z=0$ and $q_t=m$ in Eqs.~\eqref{nps-psi4}, \eqref{nps-phi22}, \eqref{nps-phi00} and \eqref{nps-phi11-lambda}.

These discussion tells us that the detection of polarizations probes the nature of gravity. If only the plus and cross modes are detected,
then gravitation is mediated by massless spin 2 field and GR is confirmed. The detection of the breathing mode in addition to the plus
and cross modes means that gravitation is mediated by massless spin 2 and spin 0 fields. If the breathing, plus, cross
and longitudinal modes are detected, then gravitation is mediated by massless spin 2 and massive spin 0 fields.
For the discussion on the detection of polarizations, please see Refs. \cite{Isi:2017fbj,DiPalma:2017qlq}.

\section{Experimental Tests}\label{sec-exps}

\subsection{Interferometers}\label{sec-ligo}

In this subsection, the response functions of the interferometers will be computed for the transverse breathing
and longitudinal polarizations following the ideas in Refs \cite{Rakhmanov:2004eh,Corda:2007hi}.
The detection of GW170104 placed an upper bound on the mass $m_g$ of the graviton \cite{Abbott:2017vtc},
\begin{equation}\label{mub17}
  m_g<m_b=7.7\times10^{-23}\,\mathrm{eV}/c^2.
\end{equation}
In obtaining this bound,the assumed dispersion relation $E^2=p^2c^2+m_g^2$,
leads to dephasing $\delta\Psi(m_g)$ of the waves relative to the phase evolution in GR, which is given by the template of the waveform predicted by GR.
The constraint on $\delta\Psi(m_g)$ from the observations was used to derive the bound on $m_g$.
Since there are only the plus and cross polarizations in GR, this bound might not be simply applied to the scalar mode.
Whether  Eq.~\eqref{mub17} can be applied to the scalar graviton is beyond the scope of the present work.
Here we assume that the upper bound $m_b$ is applicable to the scalar field and
study how the upper bound affects the detector responses to the scalar mode.
This bound potentially places severe constraint on the effect of the longitudinal polarization on the
geodesic deviation in the massive case in the high frequency band, while the transverse breathing mode can be much stronger as long as the amplitude of scalar field $\varphi$ is large enough.
Therefore, although there exits the longitudinal mode in the massive scalar-tensor theory, the detection of its effect is likely very difficult in the high frequency band.
However, in $f(R)$ gravity, the displacement in the longitudinal direction is independent of the mass and the displacements in the transverse directions become larger for smaller mass, so this mix mode can place strong constraints on $f(R)$ gravity.

In the interferometer, photons emanate from the beam splitter, are bounced back by the mirror and received by the beam splitter again.
The round-trip propagation time when the gravitational wave is present is different from that when the gravitational wave is absent.
To simplify the calculation of the response functions, the beam splitter is placed at the origin of the coordinate system.
Then the change in the round-trip propagation time comes from two effects: the change in the relative distance between the beam splitter and the mirror due to the geodesic deviation, and the distributed gravitational redshift suffered by the photon in the field of the gravitational wave \cite{Rakhmanov:2004eh}.

First, consider the response function for the longitudinal polarization.
To this end, project the light in the $z$ direction and place the mirror at $z=L$. The response function is thus given by,
\begin{equation}\label{Yl}
\begin{split}
  Y_l(f)=&\frac{m^2}{16 \pi ^3 f^3} \Big\{4 \pi  f e^{ i2 \pi  f L (v+1)}\\
  &-\frac{e^{4 i \pi  f L}}{\left(v-1\right)^3L} \Big[e^{i2 \pi  f L \left(v-1\right)}[2 \pi  f L \left(v-1\right)\\
  &+i2 ]+2 \pi  f L \left(v-1\right)-i2 \Big]\\
  &+\frac{1}{\left(v+1\right){}^3L}\Big[2 \pi  f L \left(v+1\right)+e^{i2 \pi  f L \left(v+1\right)} \\
  &\left[2 \pi  f L \left(v+1\right)+2 i\right]-i2 \Big]\Big\},
\end{split}
\end{equation}
where $f=q_t/(2\pi)$ is the frequency of gravitational waves and the propagation speed $v$ varies with the frequency.
To obtain the response function for the transverse mode, project the light in the $x$ direction and place the mirror at $x=L$.
The response function is,
\begin{equation}\label{Yb}
  Y_b(f)=e^{i2 \pi  f L}+\frac{e^{i2  \pi  f L} [2 \pi  f L-\sin (2 \pi  f L)]}{2 \pi  f L},
\end{equation}
which is independent of the mass $m$ of the scalar field.

\begin{figure}[!htp]
  \centering
  \includegraphics[width=0.4\textwidth]{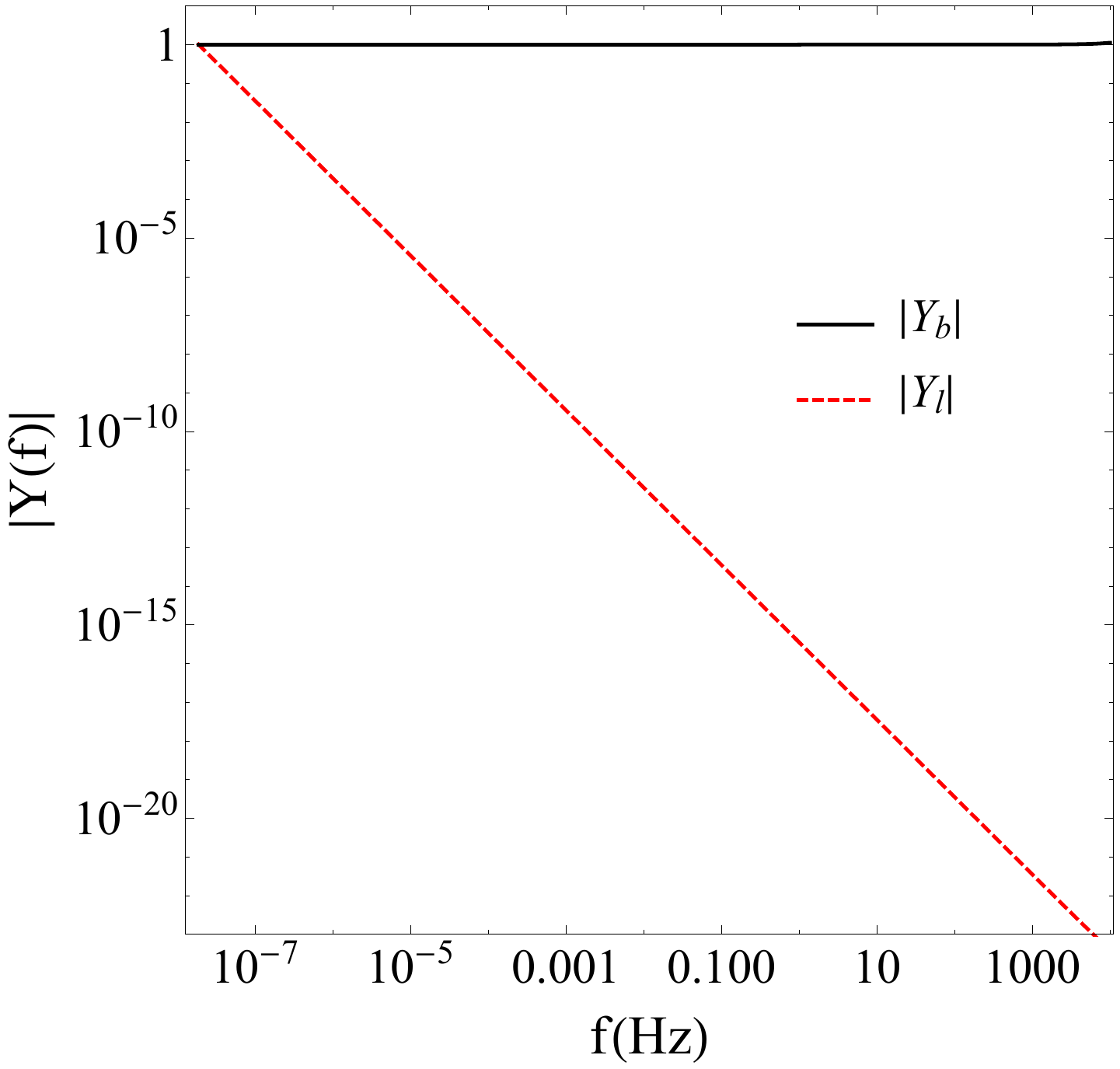}
  \caption{The absolute values of the longitudinal and transverse response functions $|Y_l(f)|$ and $|Y_b(f)|$
  as functions of $f$ for aLIGO ($L=4$ km) to a scalar gravitational wave with the mass $m=m_b$ (dashed red curve).  The solid curve denotes $|Y_b(f)|$ and the dashed curve denotes $|Y_l(f)|$.}\label{fig-yl}
\end{figure}

Figure \ref{fig-yl} shows the absolute values of the longitudinal and transverse response functions for aLIGO ($L=4$ km) to a scalar gravitational wave with the mass $m=m_b$.
Comparing the response functions shows that the transverse response is much larger than that of the longitudinal mode in high frequencies, so the detection of the longitudinal mode becomes very difficult in the high frequency band.

One can also understand the difficulty to use the interferometers to detect the longitudinal polarizations in the following way.
Table \ref{tab2} lists the magnitudes of $|\ddot x^j/x^j_0|$ normalized with $\sigma\varphi$ for the longitudinal and the transverse modes assuming the graviton mass is also $m_b$ at three different frequencies: 100 Hz, $10^{-3}$ Hz, and $10^{-7}$ Hz.
The first frequency lies in the frequency band of the ground-based detector like aLIGO and VIRGO, the second frequency lies in the frequency band of space-borne observatory like LISA, and the last frequency lies in the sensitive band of pulsar timing arrays \cite{Moore:2014lga}.

\begin{table}[h]\footnotesize
\caption{\label{tab2} The dependence of the magnitudes of $|\ddot x^j/x^j_0|$ for the longitudinal and transverse
 modes on the frequencies of gravitational waves in units of Hz$^2\sigma\varphi$ assuming the scalar mass is $m_b$.}
 \centering
 \begin{tabular}{c|c|c|c}
    \hline\hline
    & 100 Hz & $10^{-3}$ Hz & $10^{-7}$ Hz \\ \hline
    Longitudinal & 2.81$\times10^{-15}$ & 2.81$\times10^{-15}$ & 2.81$\times10^{-15}$ \\ \hline
    Transverse & $1.97\times 10^5$ & 1.97$\times10^{-5}$ & 1.97$\times10^{-13}$\\
    \hline
  \end{tabular}
\end{table}
Table \ref{tab2} shows that the effect of the longitudinal mode on the geodesic deviation is smaller than that of the transverse mode by 19 to 9 orders of  magnitude at higher frequencies.
But at the lower frequencies, i.e., at $10^{-7}$ Hz, the two modes have similar amplitudes.
Therefore, aLIGO/VIRGO and LISA might find it difficult to detect the longitudinal mode, but pulsar time arrays should be able to detect the mix polarization state with both the longitudinal and transverse modes.
The results are consistent with those for the massive graviton in a specific bimetric theory \cite{dePaula:2004bc}.

\subsection{Pulsar Timing Arrays}\label{sec-pta}

A pulsar is a rotating neutron star or a white dwarf with a very strong magnetic field. It emits a beam of the electromagnetic radiation.
When the beam points towards the Earth, the radiation can be observed, which explains the pulsed appearance of the radiation.
Millisecond pulsars can be used as stable clocks \cite{Verbiest:2009kb}. When there is no gravitational wave,
one can observe the pulses at a steady rate. The presence of the gravitational wave will alter this rate,
because it will affect the propagation time of the radiation. This will lead to a change in the time-of-arrival (TOA),
called time residual $R(t)$. Time residuals caused by the gravitational wave will be correlated between pulsars,
and the cross-correlation function is $C(\theta)=\langle R_a(t)R_b(t)\rangle$, where $\theta$ is the angular separation of pulsars $a$ and $b$,
and the brackets $\langle\,\rangle$ imply the ensemble average over the stochastic background.
This enables the detection of gravitational waves and the probe of the polarizations.

The effects of the gravitational wave in GR on the time residuals were first considered in Refs \cite{1975GReGr...6..439E,1978SvA....22...36S,Detweiler:1979wn}.
Hellings and Downs \cite{Hellings:1983fr} proposed a method to detect the effects by using the cross-correlation of the time derivative of the time residuals between pulsars,
while Jenet \textit{et.~al.} \cite{Jenet:2005pv} directly worked with the time residuals instead of the time derivative.
The later work was generalized to massless gravitational waves in alternative metric theories of gravity in Ref.~\cite{2008ApJ...685.1304L},
and further to massive gravitational waves in Refs \cite{Lee:2010cg,Lee:2014awa}.
More works have been done, for example, Refs \cite{Chamberlin:2011ev,Yunes:2013dva,Gair:2014rwa,Gair:2015hra} and references therein.

In their treatments, it is assumed that all the polarization modes have the same mass, either zero or not.
If all polarizations propagate in the $+z$ direction at the speed of light, there will be three linearly independent Killing vector fields, $\chi_1^\mu=\delta^\mu_1,\,\chi_2^\mu=\delta_2^\mu$, and $\chi_3^\mu=\delta^\mu_0+\delta^\mu_3$.
Using the conservation of $p_\mu \chi^\mu_j$ ($j=1,2,3$) for photon's 4-velocity $p^\mu$ satisfying $p^\nu\nabla_\nu p^\mu=0$, one obtains the change in the locally observed frequency of the radiation and integrates to obtain the time residual $R(t)$ \cite{1975GReGr...6..439E,Burke:1975zz,Tinto:2010hz}.
One could also directly integrate the time component of the photon geodesic equation to obtain the change in the frequency \cite{Anholm:2008wy,Lee:2010cg}.
The later method can be applied to massive case. In the present work, a different method will be used by simply calculating the 4-velocities of the photon and observers on the Earth and the pulsar.
Since different polarizations propagate at different speeds, there are not enough linearly independent Killing vector fields. The first method cannot be used any way.

\begin{figure}[h]
\centering
\includegraphics[width=0.3\textwidth]{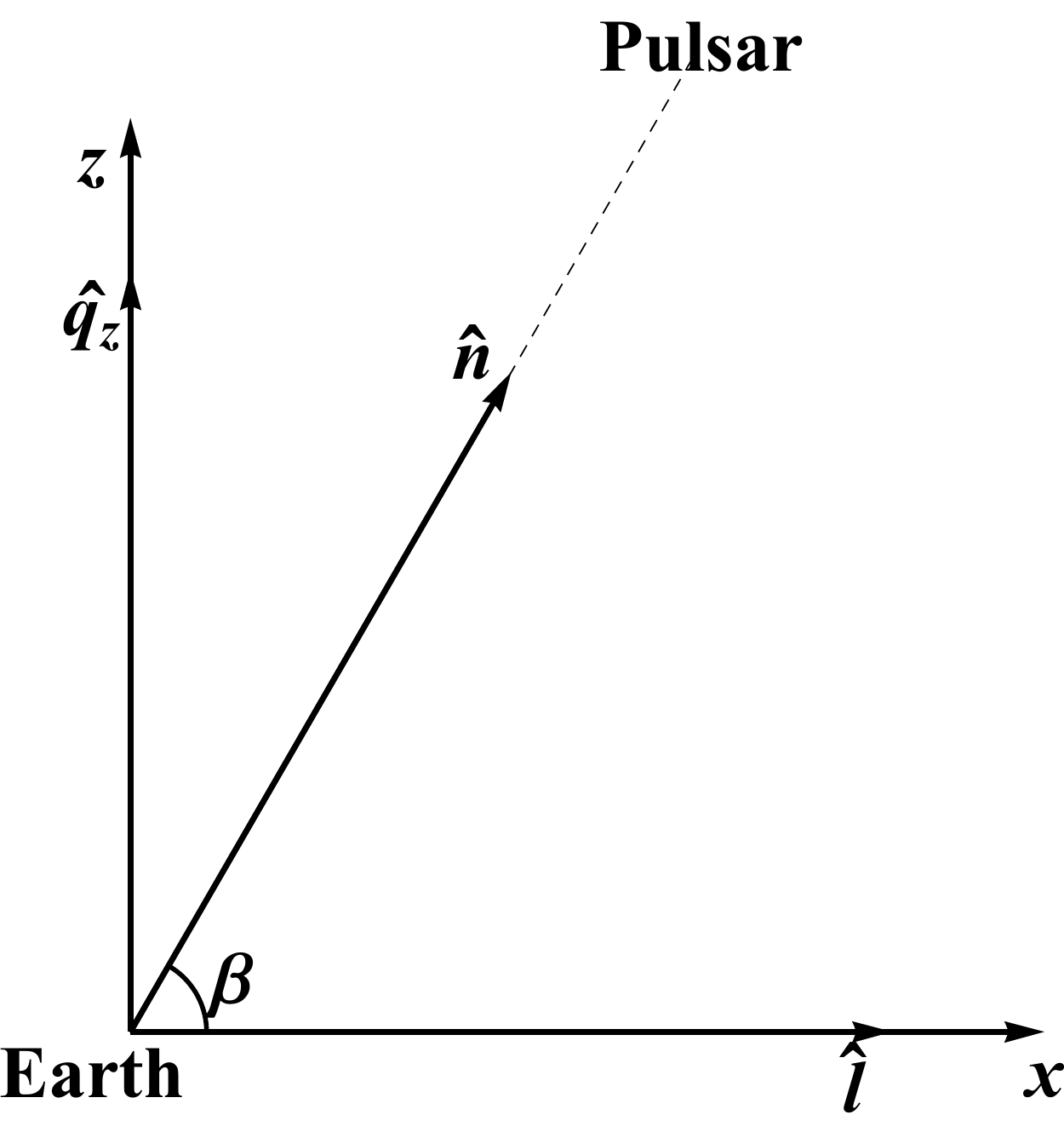}
\caption{The gravitational wave is propagating in the direction of $\hat q_z$, and the photon is traveling in $-\hat n$ direction at the leading order. $\hat l$ is perpendicular to $\hat q_z$ and in the same plane determined by $\hat q_z$ and $\hat n$. The angle between $\hat n$ and $\hat l$ is $\beta$.}\label{fig-coord}
\end{figure}

In order to calculate the time residual $R(t)$ caused by the gravitational wave solution \eqref{wavesol-h} and \eqref{wavesol-s}, one sets up a coordinate system shown in Figure \ref{fig-coord}, so that when there is no gravitational wave, the Earth is at the origin, and the distant pulsar is assumed to be stationary in the coordinate system and one can always orient the coordinate system such that the pulsar is located at $\vec x_p=(L\cos\beta,0,L\sin\beta)$.
$\hat q_z$ is the unit vector pointing to the direction of the gravitational wave, $\hat n$ is the unit vector connecting the Earth to the pulsar, and $\hat l=\hat q_z\wedge(\hat n\wedge\hat q_z)/\cos\beta=[\hat n-\hat q_z(\hat{n}\cdot\hat q_z)]/\cos\beta$ is the unit vector parallel to the $y$ axis.
In the leading order, i.e., in the absence of gravitational waves, the 4-velocity of the photon is $\underline u^\mu=\gamma_0(1,-\cos\beta,0,-\sin\beta)$ with $\gamma_0=\ud t/\ud \lambda$ a constant and $\lambda$ an arbitrary affine parameter. The perturbed photon 4-velocity is $u^\mu=\underline{u}^\mu+v^\mu$, and since $g_{\mu\nu}u^\mu u^\nu=0$, one obtains
\begin{equation}\label{eq-v0}
  v^0=\frac{\gamma_0}{2}\tilde h_{11}\cos^2\beta-v_1\cos\beta-v_3\sin\beta.
\end{equation}
Note that one chooses the gauge such that $e_{11}=-e_{22}$, $e_{12}=e_{21}$ are the only nonvanishing amplitudes for $\tilde h_{\mu\nu}$, which can always be made.
The photon geodesic equation is
\begin{equation}\label{eq-phogeo}
  \begin{split}
  0=&\frac{\ud u^\mu}{\ud \lambda}+\Gamma^\mu{}_{\rho\sigma}u^\rho u^\sigma\\
  \approx&\gamma_0\frac{\ud u^\mu}{\ud t}+\Gamma^\mu{}_{\rho\sigma}\underline u^\rho \underline u^\sigma.
  \end{split}
\end{equation}
The calculation shows that
\begin{eqnarray}
  v^0&=&\gamma_0\Big\{\sigma\varphi_0\cos[(q_t+q_z\sin\beta)t-q_z(L+t_e)\sin\beta]\nonumber\\
  &&-\frac{e_{11}}{2}(1-\sin\beta)\cos[\Omega(1+\sin\beta)t\nonumber\\
  &&-\Omega (L+t_e)\sin\beta]\Big\}, \label{eq-v0v}\\
  v_1&=&\gamma_0\{-\sigma\varphi_0\cos\beta\cos[(q_t+q_z\sin\beta)t-q_z(L+t_e)\sin\beta]\nonumber\\
  &&+e_{11}\cos\beta\cos[\Omega(1+\sin\beta)t-\Omega (L+t_e)\sin\beta]\}, \label{eq-v1v}\\
  v_2&=&\gamma_0e_{12}\cos\beta\cos\Omega[(1+\sin\beta)t-(L+t_e)\sin\beta], \label{eq-v2v}\\
  v_3&=&\gamma_0\Big\{-\sigma\varphi_0\sin\beta\cos[(q_t+q_z\sin\beta)t-q_z(L+t_e)\sin\beta]\nonumber\\
  &&-\frac{e_{11}}{2}(1-\sin\beta)\cos[\Omega(1+\sin\beta)t\nonumber\\
  &&-\Omega (L+t_e)\sin\beta]\Big\},\label{eq-v3v}
\end{eqnarray}
where $t_e$ is the time when the photon is emitted from the pulsar. Eq. (\ref{eq-v0v}) is consistent with Eq. (\ref{eq-v0}).

The 4-velocities of the Earth and the pulsar also change due to the gravitational wave. Take the 4-velocity of the Earth for instance.
Suppose when the gravitational wave is present, its 4-velocity is given by $u_e^\mu=u_e^0(1,\vec v_e)$. The normalization of $u^\mu_e$ implies that
\begin{equation}\label{u0}
  u^0_e=1+\frac{1}{2}h_{00}+O(v_e^2).
\end{equation}
The geodesic equation for the Earth is,
\begin{equation}\label{accappexp}
\begin{split}
  0=&\frac{\ud^2x^\mu}{\ud\tau^2}+\Gamma^\mu{}_{\rho\nu}\frac{\ud x^\rho}{\ud\tau}\frac{\ud x^\nu}{\ud\tau}\\
  \approx&(u^0_e)^2\left(\frac{\ud^2x^\mu}{\ud t^2}+\Gamma^\mu{}_{00}\right)+u^0_e\frac{\ud u^0_e}{\ud t}\frac{\ud x^\mu}{\ud t},
\end{split}
\end{equation}
where $\tau$ is the proper time and $x^\mu$ are the coordinates of the Earth.
One sets $x=y=0$, which is consistent with Eq.~\eqref{accappexp}. Then, one obtains the following solution
\begin{gather}\label{eq-a3-sol}
  v^3_e\approx-\frac{q_z}{2q_t}\sigma\varphi_0\cos q_t t.
\end{gather}
In addition, with Eq. \eqref{u0},
\begin{equation}\label{eq-u0}
  u^0_e=1+\frac{1}{2}\sigma\varphi_0\cos q_t t.
\end{equation}
So the 4-velocity of the Earth is approximately
\begin{equation}\label{eq-e4v}
  u^\mu_e=\left(1+\frac{1}{2}\sigma\varphi_0\cos q_t t,0,0,-\frac{q_z}{2q_t}\sigma\varphi_0\cos q_t t\right).
\end{equation}
Similarly, one can obtain the 4-velocity of the pulsar, which is
\begin{equation}\label{eq-p4v}
\begin{split}
  u^\mu_p=&\left(1+\frac{1}{2}\sigma\varphi_0\cos(q_t t-q_zL\sin\beta),0,\right.\\
  &\left.0,-\frac{q_z}{2q_t}\sigma\varphi_0\cos(q_t t-q_zL\sin\beta)\right),
\end{split}
\end{equation}
up to linear order. The form of $u^\mu_p$ can be understood, realizing that $\chi_1^\mu$ and $\chi_2^\mu$ are still the Killing vector fields.

Therefore, the frequency of the photon measured by an observer comoving with the pulsar is
\begin{equation}\label{eq-freqpul}
\begin{split}
  f_e=&-u_\mu u^\mu_p\\
  =&\gamma_0\Big[1+\frac{q_t-q_z\sin\beta}{2q_t}\sigma\varphi_0\cos(q_t t_e-q_zL\sin\beta)\\
  &-\frac{e_{11}}{2}(1-\sin\beta)\cos\Omega(t_e-L\sin\beta)\Big],
  \end{split}
\end{equation}
and the frequency measured by another observer on the Earth is
\begin{equation}\label{eq-freqear}
  \begin{split}
  f_r=&-u_\mu u^\mu_e\\
  =&\gamma_0\Big[1+\frac{q_t-q_z\sin\beta}{2q_t}\sigma\varphi_0\cos q_t(t_e+L)\\
  &-\frac{e_{11}}{2}(1-\sin\beta)\cos\Omega(t_e+L)\Big].
  \end{split}
\end{equation}
The frequency shift is thus given by
\begin{equation}\label{eq-ct}
\begin{split}
  \frac{f_e-f_r}{f_r}=&\frac{q_t-q_z\hat q_z\cdot\hat n}{2q_t}\sigma\left[\varphi(t-L,L\hat n)-\varphi(t,0)\right]\\
  &-\frac{\hat n^j\hat n^k}{2(1+\hat q_z\cdot\hat n)}\left[\tilde h_{jk}(t-L,L\hat n)-\tilde h_{jk}(t,0)\right],
  \end{split}
\end{equation}
where $t=t_e+L$ is the time when the photon arrives at the Earth at the leading order.
This expression \eqref{eq-ct} can be easily generalized to a coordinate system with an arbitrary orientation and at rest relative to the original frame.
It can be checked that in the massless limit, the change in the ratio of frequencies agrees with Eq.~(2) in Ref.~\cite{2008ApJ...685.1304L}.
The contribution of the massive scalar field also agrees with Eq.~(2) in Ref.~\cite{Lee:2014awa} which was derived provided all polarizations have the same mass.

Eq.~\eqref{eq-ct} gives the frequency shift caused by a monochromatic gravitational wave. Now, consider the contribution of a stochastic gravitational wave background which consists of monochromatic gravitational waves,
\begin{gather}
  \tilde h_{jk}(t,\vec x)=\nonumber\\
  \sum_{P=+,\times}\int_{-\infty}^{\infty}\frac{\ud \Omega}{2\pi}\int\ud^2\hat k\Big\{\tilde h^P(\Omega,\hat k)\epsilon^P_{jk}\exp[i\Omega(t-\hat k\cdot\vec x)]\Big\}, \\
  \varphi(t,\vec x)=\nonumber\\
  \int_{-\infty}^{\infty}\frac{\ud q_t}{2\pi}\int\ud^2\hat q_z\Big\{\varphi_0(q_t,\hat q_z)\exp[i(q_tt-q_z\hat q_z\cdot\vec x)]\Big\},
\end{gather}
where $\tilde h^P(\Omega,\hat k)$ is the amplitude of the gravitational wave propagating in the direction $\hat k$ at the angular frequency $\Omega$ with the polarization $P=+$ or $\times$,
$\epsilon^P_{jk}$ is the polarization tensor for the polarization $P$, and $\varphi_0(q_t,\hat q_z)$ is the amplitude for the scalar gravitational wave propagating in the direction $\hat q_z$ at the angular frequency $q_t$.
Let $\hat i,\,\hat j,\,\hat k$ form a right-handed triad frame so that $\hat i\times\hat j=\hat k$, then $\epsilon^P_{jk}$ is given by
\begin{equation}\label{polt}
  \epsilon^+_{jk}=\hat i_j\hat i_k-\hat j_j\hat j_k,\quad \epsilon^\times_{jk}=\hat i_j\hat j_k+\hat j_j\hat i_k.
\end{equation}
Assuming the gravitational wave background is isotropic, stationary and independently polarized, one defines the characteristic strains $\tilde h^P_c$ and $\varphi_c$ in the following way,
\begin{gather}
 \langle \tilde h^{P*}(\Omega,\hat k)\tilde h^{P'}(\Omega',\hat k')\rangle=\nonumber\\
 \delta(\Omega-\Omega')\delta(\hat k-\hat k')\delta^{PP'}\frac{\pi|\tilde h_c^P|^2(\Omega)}{4\Omega},\\
\langle \varphi_0^*(q_t,\hat q_z)\varphi_0(q'_t,\hat q'_z)\rangle=\delta(q_t-q'_t)\delta(\hat q_z-\hat q'_z)\frac{|\varphi_c|^2(q_t)}{q_t},\label{defvc}
\end{gather}
where the star $*$ indicates the complex conjugation.

Since the gravitational wave with the plus or cross polarization behaves exactly the same way as in GR which can be found in Ref.~\cite{2008ApJ...685.1304L},
the focus will be on the contribution of the scalar field in the following discussion.
The total time residual in TOA caused by the stochastic gravitational wave background is
\begin{equation}\label{rtsl}
  R(T)=\int_{-\infty}^{\infty}\frac{\ud q_t}{2\pi}\int\ud^2\hat q_z\int_{0}^{T}\ud t\frac{f_e-f_r}{f_r},
\end{equation}
where the argument $T$ is the total observation time. Substituting Eq.~\eqref{eq-ct} (ignoring the second line) in, one obtains
\begin{equation}\label{rtsl2}
\begin{split}
  R(T)=&\int_{-\infty}^{\infty}\frac{\ud q_t}{2\pi}\int\ud^2\hat q_z\frac{q_t-q_z\hat q_z\cdot\hat n}{i2q_t^2}\varphi_0(q_t,\hat q_z)(e^{iq_tT}-1)\times\\
  &[1-e^{-i(q_t+q_z\hat q_z\cdot n)L}].
  \end{split}
\end{equation}
Therefore, one can now consider the correlation between two pulsars $a$ and $b$ which are located at positions $\vec x_a=L_1\hat n_1$ and $\vec x_b=L_2\hat n_2$, respectively.
The angular separation is $\theta=\arccos(\hat n_1\cdot \hat n_2)$. The cross-correlation function is thus given by
\begin{equation}\label{cxcorel}
\begin{split}
  C(\theta)=&\langle R_a(T)R_b(T)\rangle\\
   =&\int_{m}^{\infty}\ud q_t\int\ud^2\hat q_z\frac{(q_t-q_z\hat q_z\cdot\hat n_1)(q_t-q_z\hat q_z\cdot\hat n_2)}{4\pi^2 q_t^5}\times\\
   &|\varphi_c|^2(q_t)\mathcal{P},
\end{split}
\end{equation}
where
 \begin{equation}\label{eq-def-p}
   \mathcal{P}=1-\cos\Delta_1-\cos\Delta_2+\cos(\Delta_1-\Delta_2),
 \end{equation}
and $\Delta_j=(q_t+q_z\hat q_z\cdot \hat n_j)L_j$ with $j=1,2$.
In obtaining this result, one uses Eq.~\eqref{defvc}, and takes the real part.
In addition, $T$ drops out, since the ensemble average also implies the average over the time \cite{2008ApJ...685.1304L}.

Since the gravitational wave background is assumed to be isotropic, one can calculate $C(\theta)$ by setting
\begin{gather}\label{defns}
  \hat n_1=(0,0,1), \\
  \hat n_2=(\sin\theta,0,\cos\theta).
\end{gather}
Also, let $\hat q_z=(\sin\theta_g\cos\phi_g,\sin\theta_g\sin\phi_g,\cos\theta_g)$, so
\begin{gather}\label{deltas}
  \Delta_1=(q_t+q_z\cos\theta_g)L_1,\\
  \Delta_2=[q_t+q_z(\sin\theta_g\cos\phi_g\sin\theta+\cos\theta_g\cos\theta)]L_2,
\end{gather}
and
\begin{equation}\label{cxcor1}
\begin{split}
  C(\theta)=&\int_{m}^{\infty}\ud q_t\int\ud\theta_g\ud\phi_g\sin\theta_g\frac{|\varphi_c|^2(q_t)}{4\pi^2q_t^5}(q_t-q_z\cos\theta_g)\\
  &[q_t-q_z(\sin\theta_g\cos\phi_g\sin\theta+\cos\theta_g\cos\theta)]\mathcal{P}.
\end{split}
\end{equation}

In the observation,  pulsars are far away enough, so that $q_tL_j\gg1$.
This implies that one can approximate $\mathcal{P}\approx1$ when $\theta\ne0$, as the phases in cosines in the definition \eqref{eq-def-p} of $\mathcal{P}$  oscillate fast enough.
The integration can be partially done, yielding
\begin{equation}\label{cxcor2}
  C(\theta)\approx\int_{m}^{\infty}\ud q_t\frac{|\varphi_c|^2(q_t)}{\pi q_t^3}\left(1+\frac{q_z^2}{3q_t^2}\cos\theta\right).
\end{equation}
But for $\theta=0$, one cannot simply set cosines in $\mathcal{P}$ to 1. In this case, one actually considers the auto-correlation function, so $\hat n_1=\hat n_2$ and $L_1=L_2=L$.
The auto-correlation function is thus given by
\begin{equation}\label{autocor}
\begin{split}
  C(0)=&2\int_{m}^{\infty}\ud q_t\frac{|\varphi_c|^2(q_t)}{\pi q_t^3}\left[1+\frac{q_z^2}{3q_t^2}+\frac{2}{q_t^2L^2}\cos q_tL\cos q_zL\right.\\
  &\left.+\left(\frac{2}{q_tL}-\frac{1}{q_zL}+\frac{2}{q_t^2q_zL^3}-\frac{q_z}{q_t^2L}\right)\cos q_tL\sin q_zL\right.\\
  &\left.-\frac{2}{q_tq_zL^2}\sin q_tL\sin q_zL\right].
\end{split}
\end{equation}
Note that the observation time $T$ sets a natural cutoff for the frequency, i.e., $q_t\ge 2\pi/T$. So the lower integration limits in Eqs.~\eqref{cxcor2} and \eqref{autocor} should be replaced by $\text{Min}\{m,2\pi/T\}$.

\begin{figure}
  \centering
  \includegraphics[width=0.4\textwidth]{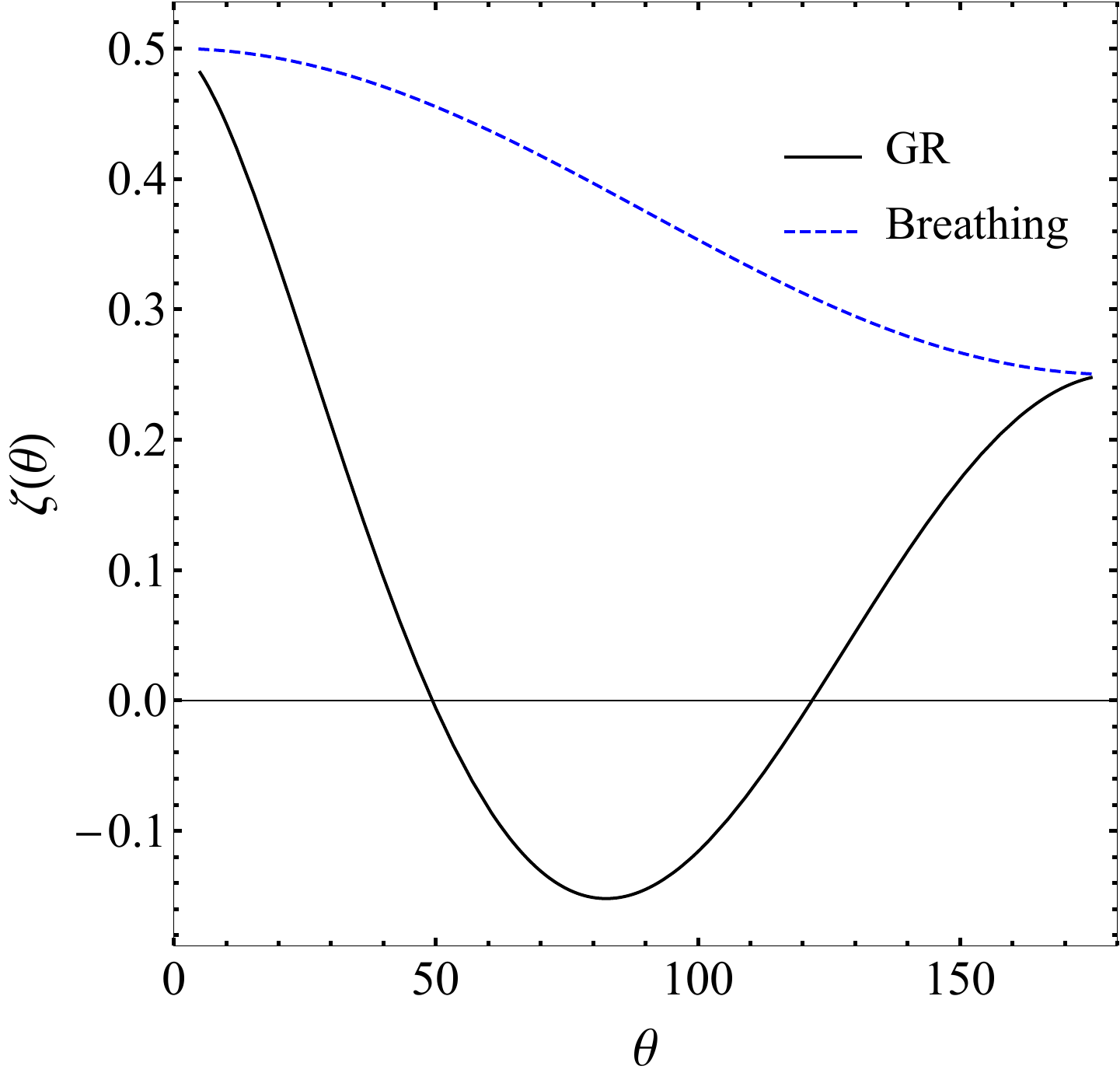}
  \includegraphics[width=0.4\textwidth]{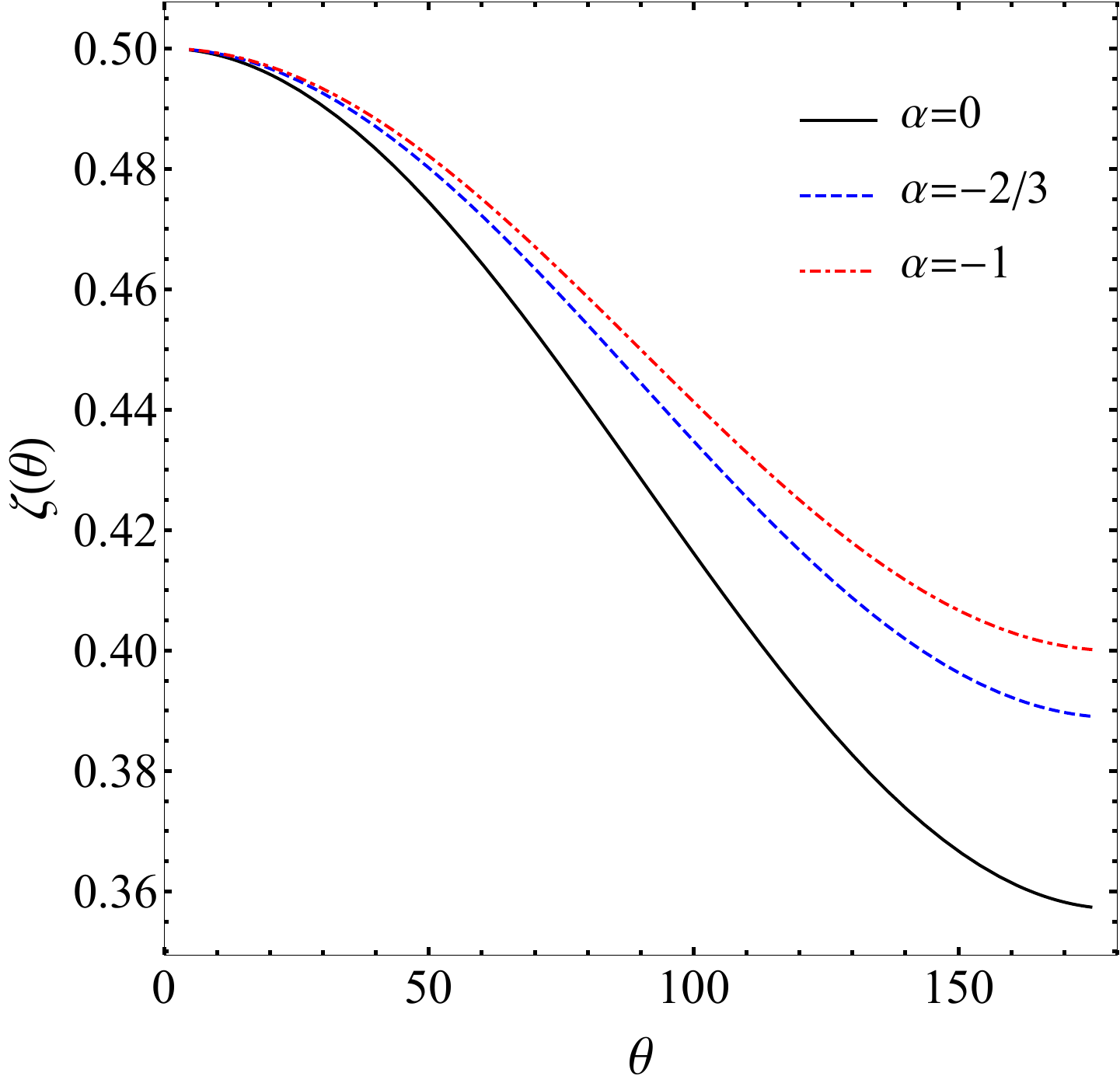}
  \caption{The normalized cross-correlation functions $\zeta(\theta)=C(\theta)/C(0)$. The upper panel shows the cross-correlations when the scalar field is massless, i.e.,
  when there is no longitudinal polarization. The solid curve is for familiar GR polarizations (i.e., the plus or cross ones),
  and the dashed curve for the transverse breathing polarization.
  The lower panel shows the cross-correlations induced together by the transverse breathing and longitudinal polarizations when the mass
  of the scalar field is taken to be $m_b=7.7\times10^{-23}\,\mathrm{eV}/c^2$. The calculation was done assuming $T=5$ yrs.}\label{fig-normcorr}
\end{figure}

Usually, one assumes that $\varphi_c(q_t)$ takes a form of  $\varphi_c(q_t)\propto(q_t/q_t^c)^\alpha$ with $q_t^c$ some characteristic angular frequency.
$\alpha$ is called the power-law index, and usually, $\alpha=0,\,-2/3$ or $-1$ \cite{2008ApJ...685.1304L,Romano:2016dpx}.
One can numerically integrate Eqs.~\eqref{cxcor2} and \eqref{autocor} to obtain the so-called normalized correlation function $\zeta(\theta)=C(\theta)/C(0)$.
In the integration, suppose the observation time $T=5$ years and the mass of the scalar field is $m_b=7.7\times10^{-23}$ eV$/c^2$ \cite{Abbott:2017vtc}.
The distance $L$ takes a large enough value so that $q_tL\gg1$ is satisfied. This gives rise to the right panel in Figure \ref{fig-normcorr},
where the power-law index $\alpha$ takes different values.
Together shown are the normalized correlation functions for the plus and cross polarizations (labeled by ``GR"),
and for the transverse breathing polarization (labeled by ``Breathing") in the left panel.
The normalized correlation function for the transverse breathing polarization can be obtained simply by setting the mass $m=0$.
It is clear that $\zeta(\theta)$ behaves very differently for $\tilde h_{\mu\nu}$ and $\varphi$,
so it is possible to determine whether there are polarizations induced by the scalar field using pulsar timing arrays.
Note that since $L$ is large enough, $\zeta(\theta)$ barely changes with $L$.

\begin{figure}
  \centering
  \includegraphics[width=0.45\textwidth]{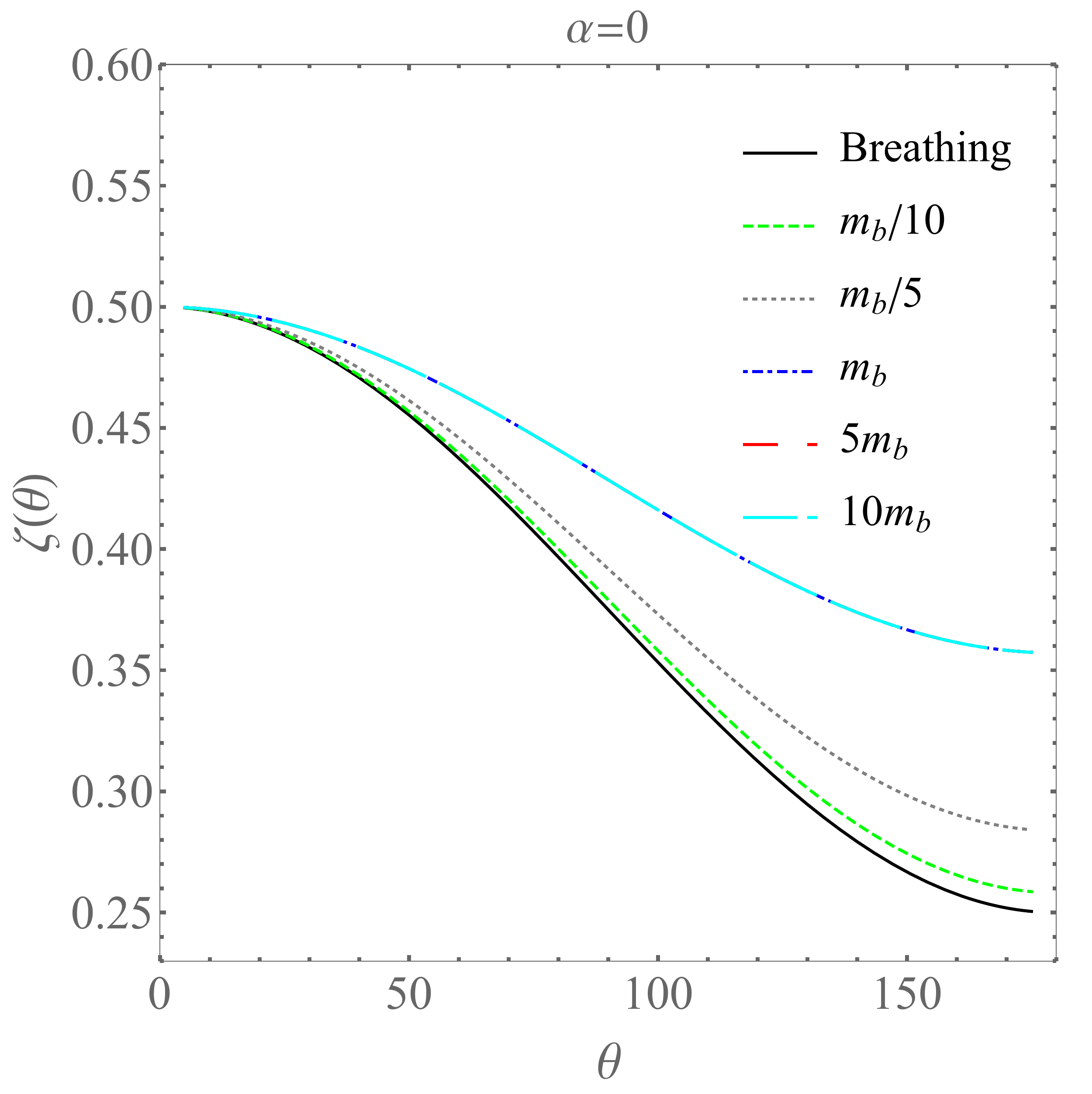}
  \caption{$\zeta(\theta)$ as a function of scalar mass $m$ at $\alpha=0$. The calculations were also done for a 5-year observation.}\label{fig-alpha0}
\end{figure}

One could also vary the mass of the scalar field.
The result is shown in Figure \ref{fig-alpha0}. This figure displays $\zeta(\theta)$ for six different multiples of $m_b$, including $m=0$ labeled by ``Breathing".
The power-law index $\alpha$ is chosen to be 0.
It shows that at large angles, $\zeta(\theta)$ is very sensitive to smaller masses with $m_b/5\lesssim m\lesssim m_b$, but for $m\gtrsim m_b$ or $m\lesssim m_b/5$, $\zeta(\theta)$ almost remains the same.
Therefore, the measurement of the correlation can be used to constrain the mass of the scalar field.

In Ref.~\cite{Lee:2014awa}, Lee also analyzed the time residual of TOA caused by massive gravitational waves and calculated the cross-correlation functions.
His results (the right two panels in his Figure 1) differ from those on the right panel in Figure~\ref{fig-normcorr}, because in his treatment, the longitudinal and the transverse polarizations were assumed to be independent.
In Horndeski theory, however, it is not allowed to calculate the cross-correlation function separately for the longitudinal and the transverse polarizations, as they are both excited by the same field $\varphi$ and the polarization state is a single mode.

\section{Conclusion}\label{sec-con}

This work analyzes the gravitational wave polarizations in the most general scalar-tensor theory of gravity,  Horndeski theory.
It reveals that there are three independent polarization modes: the mixture state of the transverse breathing $\hat{P}_b=R_{txtx}+R_{tyty}$
and longitudinal $\hat{P}_l=R_{tztz}$ polarizations for the massive scalar field,
and the usual plus $\hat{P}_+=-R_{txtx}+R_{tyty}$ and cross $\hat{P}_\times=R_{txty}$ polarizations for the massless gravitons.
These results are consistent with the three propagating degrees of freedom in Horndeski theory.
Since the propagation speed of the massive gravitational wave depends on the frequency and is smaller than the speed of light,
the massive mode will arrive at the detector later than the massless gravitons. In addition to the difference of the propagation speed,
the presence of both the longitudinal and breathing states without the vector-$x$ and vector-$y$ states
are also the distinct signature of massive scalar degree of freedom for graviton.

Using the NP variables,
we find that $\Psi_2=0$. For null gravitational waves, this means that the longitudinal mode does not exist. However,
our results show that the longitudinal mode exists in the massive case even though $\Psi_2=0$. We also find that the NP variable $\Phi_{00}\neq 0$,
and $\Phi_{00}$, $\Phi_{11}$ and $\Lambda$ are all proportional to $\Phi_{22}$. These results are in conflict with those for massless gravitational waves in \cite{Eardley:1974nw},
so the results further support the conclusion that the classification of the six polarizations for null gravitational waves derived from the little group $E(2)$ of
the Lorentz group is not applicable to the massive case. Although the longitudinal mode exists for the massive scalar field,
it is difficult to be detected in the high frequency band because of the suppression by the extremely small graviton mass upper bound.
Compared with aLIGO/VIRGO and LISA, pulsar timing arrays might be the primary tool to detect both the transverse breathing and longitudinal modes due to the massive scalar field.
In the massless case, the longitudinal mode disappears and the mix state reduces to the pure transverse breathing mode.
Brans-Dicke theory and $f(R)$ gravity are subclasses of Horndeski theory, so the general results obtained can be applied to those theories.
Despite the fact that in $f(R)$ gravity,
the longitudinal mode does not depend on the mass of graviton, its magnitude is much smaller than the transverse one in the high frequency band,
which makes its detection unlikely by the network of aLIGO/VIGO and LISA, too.
It is interesting that the transverse mode becomes stronger for smaller graviton mass in the $f(R)$ gravity,
so the detection of the mixture state can place strong constraint on $f(R)$ gravity.

For null gravitational waves, the presence of the longitudinal mode means that all six polarizations can be detected in some coordinate systems.
For Horndeski theory, we find that the vector modes are absent even though the longitudinal mode is present.
Since the massive scalar field excites both the breathing and longitudinal polarizations,
while the massless scalar field excites the breathing mode only,
the detection of polarizations can be used to understand the nature of gravity.
If only the plus and cross modes are detected,
then gravitation is mediated by massless spin 2 field and GR is confirmed. The detection of the breathing mode in addition to the plus
and cross modes means that gravitation is mediated by massless spin 2 and spin 0 fields. If the breathing, plus, cross
and longitudinal modes are detected, then gravitation is mediated by massless spin 2 and massive spin 0 fields.

\begin{acknowledgement}
We would like to thank Ke Jia Lee for useful discussions. This research was supported in part by the Major Program of the National Natural Science Foundation of China under Grant No. 11690021 and the National Natural Science Foundation of China under Grant No. 11475065.
\end{acknowledgement}


\end{document}